\documentclass[10pt,letterpaper,twocolumn]{article}
\usepackage[english]{babel}
\usepackage{graphicx}

\newcommand{\alt}{\mathbin{\lower 3pt\hbox
   {$\rlap{\raise 5pt\hbox{$\char'074$}}\mathchar"7218$}}}
\newcommand{\agt}{\mathbin{\lower 3pt\hbox
   {$\rlap{\raise 5pt\hbox{$\char'076$}}\mathchar"7218$}}}

\begin{document}

\setcounter{footnote}{0}
\setcounter{equation}{0}
\setcounter{figure}{0}
\setcounter{table}{0}

\title{\large\bf Mutual distribution of two partial
solutions \\ in 1D localization:
new information on the phase transition }

\author{\small  I. M. Suslov \\
\small P.L.Kapitza Institute for Physical Problems,
119334 Moscow, Russia \\
\small E-mail: suslov@kapitza.ras.ru\\
{}\\
\parbox{150mm}{\footnotesize \,We consider the mutual distribution of two
linearly independent solutions $y_1(x)$ and $y_2(x)$ of the 1D
Schr${\rm\ddot o}$dinger equation with a random potential. Since
individual distributions of $y_1$ and $y_2$ are log-normal, it is naturally
to suggest that their mutual distribution is also log-normal. Such hypothesis
is confirmed in the deep of the allowed and forbidden bands, but failed near
the initial band edge. The mechanism of deviations from the log-normal form
is elucidated, and the first correction to it is calculated. The latter allows
to demonstrate broadening of the spectral lines in the universal conductance
fluctuations. A lot of new information is obtained on the phase transition in
the distribution $P(\psi)$, where $\psi$ is a combined phase entering the
evolution equations. According to the previous publications, this transition
is related with appearance of the imaginary part of $\psi$ at a certain energy
${\cal E}_0$, and is not accompanied by singularities in the system resistance.
The real sense of this transition consists in the change of configuration of
four Lyapunov exponents, which determine the general solution: there are two
pairs of complex-conjugated exponents for ${\cal E}>{\cal E}_0$,  while for
${\cal E}<{\cal E}_0$ all exponents become real. Realization of two different
configurations is confirmed for energies in the deep of the allowed and
forbidden bands; it proves the existence of the singular point ${\cal E}_0$
at the formal level. The phase transition can be observed in optical systems,
tracing the sign of the field in a wave, when the coordinate is changed. } }

\date{}
\maketitle


\setcounter{footnote}{0}
\setcounter{equation}{0}
\setcounter{figure}{0}
\setcounter{table}{0}


\begin{center}
{\bf 1. Introduction}
\end{center}

The localization theory originates from the papers by Anderson
\cite{0} and Mott \cite{1,2}, received a new life with
incorporation of scaling ideas \cite{2a}, and now it is actively
discussed in the context of many-body localization
\cite{500}--\cite{508}.  In the present paper we consider the
 mutual distribution of two linearly independent solutions  of
the 1D Schr${\rm\ddot o}$dinger equation with a random potential.
The physical motivation for it is three-fold:
(a) the knowledge of such distribution allows to
determine the discrete frequencies in universal conductance
fluctuations (Secs.\,1,\,10); (b) it provides new information on
the unusual phase transition,
discussed previously (Secs.\,1,\,5);
(c) it gives new insight in the transfer matrix
approach used for numerical  estimation of the critical behavior
for the Anderson transition (Sec.11).

It is well-known, that a solution of the 1D
Schr${\rm\ddot o}$dinger
equation in the forbidden band of an ideal crystal is given
by a superposition of the growing and decreasing exponents
$$
y_1(x)={\rm e}^{\kappa x}\,,\qquad
y_2(x)={\rm e}^{-\kappa x}\,.
\eqno(1)
$$
As was indicated firstly by Mott \cite{1,2}, in the
disordered systems a situation typical for the forbidden band
remains qualitatively valid for all energies. Indeed, let a wave
of the unit amplitude falls from the left on the system of point
scatterers located in the interval $(0,L)$ (Fig.1). In the case
of identical and periodically arranged scatterers the system
behaves as an effective transparent media, and the transmission
amplitude $t$ oscillates with a change of $L$, but always remains
of the order of unity. If a disorder is present in the system,
then the transmission coefficient  $|t|^2$ decays exponentially
with  $L$, as known practically for the transmission
of waves through the layer of semi-transparent material; for
large $L$,  the damping decrement
appears to be a well-defined (deterministic) quantity. It
indicates the exponential decay  of the
incident wave inside the system, with the weakly fluctuating
decrement $\kappa$.  If a wave is incident from the right to
left, then its attenuation corresponds to
existence of the increasing solution (from the left to right)
with the same exponent $\kappa$.
According to Mott, it indicates
localization of all states in  1D disordered systems, since the
only possibility to obtain the wave function, restricted in the
whole space, consists in integration of the  Schr${\rm\ddot o}$dinger
equation from two ends of the system and matching at
some point inside of it.

Since the 1D Schr${\rm\ddot o}$dinger equation has a
structure
$$
y''+f(x)y=0 \,,
\eqno(2)
$$
its Wronskian accepts a constant value \cite{3}
$$
y'_1 y_2-y_1 y'_2={\rm const}\,.
\eqno(3)
$$
If one suggest that
$$
y_1(x)={\rm e}^{\kappa_1 x}\,,\qquad
y_2(x)={\rm e}^{\kappa_2 x}\,
\eqno(4)
$$
with weakly fluctuating exponents $\kappa_i$, then it is
easy to obtain
$$
\kappa_1=-\kappa_2 \,,
\eqno(5)
$$
which returns to Eq.1, but with the generally complex-valued
parameter $\kappa$. The latter is clear from the
fact, that $\kappa$ is pure imaginary inside the allowed band of
the pure material, $\kappa=i k$, where $k$ accepts large values
in the deep of the band; appearing of weak disorder leads to
arising of the small real part of $\kappa$, but cannot eliminate
its large imaginary part.

\begin{figure}
\centerline{\includegraphics[width=3.2 in]{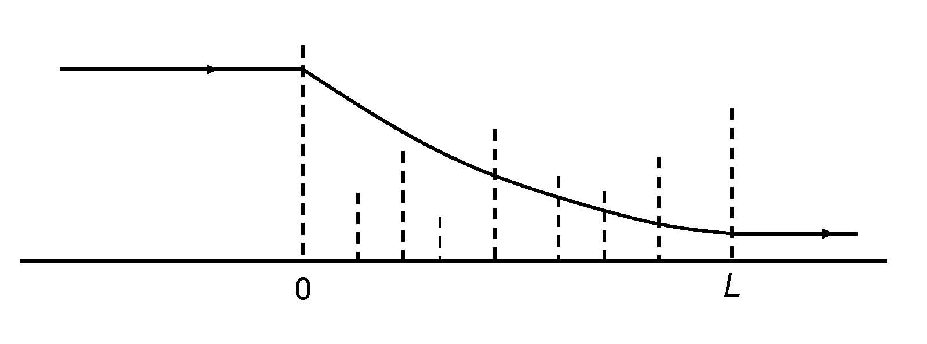}}
\caption{ \small  If a wave of the unit amplitude falls on
the system of point scatterers, then its amplitude exponentially
decreases in the depth of system, if a disorder is present in
it. } \label{fig1}
\end{figure}

The Mott argumentation looks rather convincing, if
exponents $\kappa_1$ and $\kappa_2$ are weakly fluctuating, i.e.
self-averaging. The latter property for a real part of
$\kappa_i$ follows from the Oceledets theorem \cite{4} and leads
to the notion of the Lyapunov exponents. Let introduce the
transfer matrix $T$ in the wave representation, relating the
amplitudes of waves on the left ($Ae^{ikx}+Be^{-ikx}$) and on
the right ($Ce^{ikx}+De^{-ikx}$) of a scatterer,
$$
\left ( \begin{array}{cc} C \\ D \end{array}\right)
=  T \left ( \begin{array}{cc} A \\ B \end{array} \right)\,
\,,
\eqno(6)
$$
which can be parametrized in the form
$$
 T= \left ( \begin{array}{cc} \!\!\! 1/t\! &\! - r/t \!\!\\
\!\!- r^*/t^* \!&\! 1/t^* \!\!\!\end{array} \right)\,
= \left ( \begin{array}{cc}
\!\!\sqrt{\rho\!+\!1}\, e^{i\varphi}\!\! &
\!\!\sqrt{\rho} \,e^{i\theta}\!\!
\\ \!\!\sqrt{\rho}\, e^{-i\theta}\!\!
&\!\! \sqrt{\rho\!+\!1}\,
e^{-i\varphi}\!\! \end{array} \right)\,,
\eqno(7)
$$
if a time-reversal invariance is suggested \cite{5}:
here $t$ and $r$ are the transmission and reflection
amplitudes, and $\rho=|r/t|^2$ is the dimensionless
Landauer resistance \cite{6}. For a successive arrangement of
scatterers their transfer-matrices are multiplied, so the
matrix $T$ for the whole system is represented as a product of
$n$ random matrices\,\footnote{\,Strictly speaking
\cite{13,126}, the matrices $T_i$ are given by  the products
$T_{\epsilon_i} T_{\delta_i}$, where $T_{\epsilon_i}$
correspond to the point scatterers, while  $T_{\delta_i}$
describe the intervals between scatterers.}:
$$
T=T_1 T_2 \ldots T_n \,.
\eqno(8)
$$
Let attempt to represent this product in the form $T_0^n$,
where $T_0$ is a constant matrix; in fact, it is possible for its
Hermitian part. The matrix $T$, as any matrix, allows
factorization $T=HU$, where $H$ and $U$ are the
Hermitian and unitary matrices; the latter conserves the norm of
a vector and responds for oscillations, while the former controls
a systematic growth or decrease. Let define the matrix $T_0$ as
$$
T_0=(TT^+)^{1/2n}=(HH^+)^{1/2n}=H^{1/n} \,.
\eqno(9)
$$
According to the Oceledets theorem \cite{4}, the
eigenvalues of this Hermitian matrix can be written
in the form $\lambda_{1,2}={\rm e}^{\pm \kappa}$, where
$\kappa$ and $-\kappa$ are the characteristic Lyapunov
exponents; in the limit $n\to\infty$ they tends to
deterministic (non-random) values. Applying the matrix
$T$ to the constant column and ignoring oscillations, one comes
to the partial solutions $e^{\pm \kappa n}$, analogous to
(1).\,\footnote{\,For definiteness, we have in mind the 1D
Anderson model, which contains a scatterer at each lattice site,
so $n$ corresponds to the coordinate $x$ in units of the lattice
spacing $a_0$. Near the initial band edge the Anderson model
describes adequately practically any random potential with
short-range correlations  (Sec.7). }

In fact, the physical papers \cite{7}--\cite{13} provide
essentially stronger assertions.  It is easy to verify,
that eigenvalues of the Hermitian matrix
$$
T T^+= \left ( \begin{array}{cc}
\!\!1+2\rho\!\,  &
\!\!2\sqrt{\rho(1+\rho)} \,e^{i\theta+i\varphi}\!\!
\\
\!\!2\sqrt{\rho(1+\rho)} \,e^{-i\theta-i\varphi}\!\!
&\!\! 1+2\rho\!\,\!\! \end{array} \right)\,
\eqno(10)
$$
can be represented in the form
$$
\lambda_{1,2}={\rm e}^{\pm z}\,, \qquad
{\rm ch}{z}=1+2\rho\,\qquad(z>0).
\eqno(11)
$$
In the large $n$ limit, the typical values of $\rho$ are large
and $z\approx \ln \rho$, while $\rho$ obeys the log-normal
distribution \cite{7}--\cite{13}; it leads to the Gaussian
distribution for $z$,
$$
P(z)=\frac{1}{ \sqrt{4\pi D n}}
\exp\left\{-\frac{(z-vn)^2}{4Dn}\right\}\,,
\eqno(12)
$$
with parameters $v$ and $D$, depending on the energy ${\cal E}$
(Fig.2), counted from the lower edge of the initial band.
The mean and the variance of $z$ grows proportionally to $n$,
so the quantity $\kappa=z/n$ tends to the constant value, with
its fluctuations diminishing as  $n^{-1/2}$. The analogous
distribution for the decreasing Lyapunov exponent follows from
(12) by substitution of $-z$ for $z$.

One can see, that individual distributions of $y_1$ and $y_2$
are log-normal. It is natural to expect, that their
mutual distribution is also log-normal; namely,
if we accept
$$
y_1={\rm e}^{z_1}\,, \qquad y_2={\rm e}^{z_2}\,,
\eqno(13)
$$
then the mutual Gaussian distribution is expected for $z_1$
and $z_2$. Such hypothesis is confirmed in the deep of the
allowed and forbidden bands, but failed near the initial
band edge (Secs.\,2,\,4). Arising situation looks rather strange.
If there are serious grounds for  validity of the Gaussian
distribution (like the central limit theorem), then why it is not
valid for all energies?  If there are no such grounds, why it
is valid anywhere? The mechanism of deviations from the Gaussian
form is elucidated
in Sec.6, while its consequences are discussed in Sec.7.

Another problem with the Gaussian distribution consists in
the fact, that average values
$$
\left\langle z_1 \right\rangle = v_1 n\,, \qquad
\left\langle z_2 \right\rangle = v_2 n
\eqno(14)
$$
do not satisfy the condition $v_1=-v_2$, evident from the
previous discussion. This problem is stated more explicitly in
Sec.2 and resolved in the subsequent sections.

The interest to the distribution $P(y_1,y_2)$ is
clear from following considerations. Let take
the linear combination of two solutions (1) in the
forbidden band of the ideal system, and squaring it, come to a
superposition of exponents with parameters
$$
2\kappa\,, \qquad 0\,,  \qquad
-2\kappa \,,
\eqno(15)
$$
while raising it to the fourth power leads to a set of values
$$
4\kappa\,, \qquad 2\kappa\,, \qquad 0\,,
\qquad -2\kappa \,, \qquad -4\kappa \,.
\eqno(16)
$$
If weak disorder is introduced to the system, these exponents
change slightly and correspond to  behavior of the second and
fourth moments. According to \cite{13}, these sets of parameters
for the 1D Anderson model are determined by the roots of
algebraic equations of the third and fifth power
correspondingly. In the deep of the forbidden band one
has results
$$
2\delta+\epsilon^2\,, \quad -2\epsilon^2\,,  \quad
-2\delta+\epsilon^2
\eqno(17)
$$
for the second moments, and
$$
4\delta+6\epsilon^2, \quad
2\delta-3\epsilon^2, \quad -6\epsilon^2,  \quad
-2\delta-3\epsilon^2 , \quad -4\delta+6\epsilon^2
\eqno(18)
$$
for the fourth moments. Here $\delta=\kappa a_0$,
$\epsilon^2=W^2/(2\kappa a_0)^2$, where $W$  is an amplitude of
the random potential; the results (15), (16) correspond
to the exponents of type $\exp(\kappa x)$, while (17), (18)
to the exponents $\exp(\tilde\kappa n)$, and differ by a factor
 $a_0$.  It is easy to understand that parameters (17),
(18) can be associated with averages
$\left\langle y_1^{m_1} y_2^{m_2} \right\rangle$ with
$m_1\!+\!m_2=2$ and $m_1\!+\!m_2=4$
correspondingly. If the mutual distribution  $P(y_1,y_2)$
is known and  determined by a small number of
parameters, then one is able to establish the complete set of
exponents of type (17), (18) for all moments and all energies. In
principle, such exponents are observable. Transition to the
allowed band is produced by replacement $\delta\to i\delta$,
$\epsilon\to i\epsilon$, and the imaginary parts of the exponents
correspond to discrete frequencies of oscillations in the moments
$\left\langle \rho^m \right\rangle$, which lead to universal
conductance fluctuations \cite{301}--\cite{315}, and can be
extracted from experiment by the spectral analysis
\cite{122,123}. The real parts of the exponents can be also
extracted  \cite{123}.

Another group of questions is related with the oscillatory
behavior of solutions. As was indicated above, parameters
$\kappa_1$ and $\kappa_2$ are generally complex-valued.
Self-averaging of their real part and equality
${\rm Re}\,\kappa_1=-{\rm Re}\,\kappa_2$ follows from the
Oceledets theorem \cite{4}, while in respect of
imaginary parts the question remains open: its clarification
is one of the purposes of the paper. In the presence of
time-reversal invariance, solutions of the Schr${\rm\ddot
o}$dinger equation can be chosen real.
It requires existence of exponents
$\kappa\pm ik$, $-\kappa\pm ik$ and
representation of the general solution as a superposition of four
exponential functions\,\footnote{\,Formula (19) is somewhat
conditional due to the fact, that at small length scales
parameters  $\kappa$ and $k$ are strongly fluctuating, and their
fluctuations become weak only after averaging over small
scales.  In the limit of large concentration of weak scatterers
(Sec.7) small length scales tends to zero, and this reservation
becomes unnecessary.  Complete jusification of Eq.19 will be
given in Secs.3,\,4.  }
$$
y(x)=C_1 {\rm e}^{\kappa x+i k x}+C_2 {\rm e}^{\kappa x-i k x}
+C_3 {\rm e}^{-\kappa x+i k x}+C_4 {\rm e}^{-\kappa x-i k x}
\,.\eqno(19)
$$
One can wonder, how it agree with existence of only two
linearly independent solutions for equation (2).
One can also worry,
that for two partial
solutions (19) with different sets of $C_i$
the Wronskian does not accept a constant value.
Resolution of these questions is given in Sec.3.

\begin{figure}
\centerline{\includegraphics[width=2.4 in]{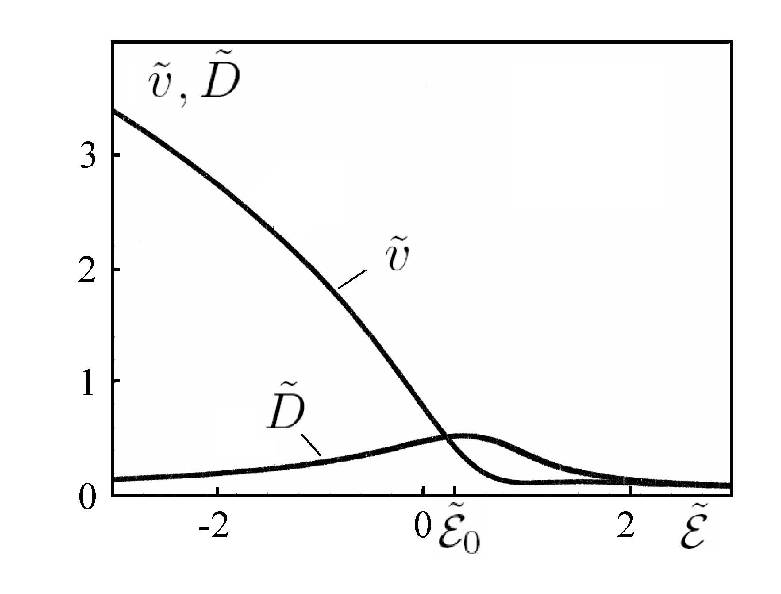}}
\caption{\small
 Dependence of parameters  $\tilde v=v/W^{2/3}$ and $\tilde
 D=D/W^{2/3}$ on the reduced energy $\tilde{\cal E}={\cal
E}/W^{4/3}$, obtained from the analysis of moments for the
transfer matrix elements \cite{13}. These moments are regular
functions of energy, which leads to regularity of the presented
dependencies. The point $\tilde{\cal E}_0$ corresponds to
the phase transition in the distribution $P(\psi)$.  }
\label{fig2}
\end{figure}

According to \cite{124}--\cite{126}, the most general evolution
equation is derived for the mutual distribution
$P(\rho,\psi,\chi)$ of the Landauer resistance $\rho$ and two
combined phases
$$
\psi=\theta-\varphi\,, \quad \chi=\theta+\varphi\,,
\eqno(20)
$$
which are directly observable in optical systems \cite{124,125}.
For large $n$ the distribution is factorized,
$P(\rho,\psi,\chi)=P(\rho)\,P(\psi)\,P(\chi)$, providing
the existence of the stationary distribution $P(\psi)$,
which determines the coefficients in the evolution
equation for $P(\rho)$. According to Sec.9, the mutual
distribution of  $y_1$ and $y_2$ is formally expressed
through $P(\rho,\psi,\chi)$,  but the practical use of this
representation is rather problematic. However, the first
correction to the Gaussian distribution can be calculated
explicitly (Sec.8). As its practical application, we
demonstrate the broadening of spectral lines in the universal
conductance fluctuations (Sec.10).

According to \cite{125,126}, at a certain energy ${\cal E}_0$ the
phase $\psi$ accepts the imaginary part, and the transfer matrix
$T$, relating the amplitudes of the running waves, transforms to
the pseudo transfer matrix \cite{13}, relating the coefficients
of growing and decreasing exponents.
The Landauer resistance $\rho$ has no singularity at the point
${\cal E}_0$, and the indicated phase transition looks
unobservable in electronic systems; its observability in
optics was justified in \cite{124,125}.  As shown below,
a real sense of this transition consists in the change of
configuration of four Lyapunov exponents in Eq.19: at the point
${\cal E}_0$, the quantity $i\langle k\rangle$ changes to
$\langle \kappa_1\rangle$,  and instead of two pairs of
complex-conjugated exponents for ${\cal E}>{\cal E}_0$,  one has
four real exponents for ${\cal E}<{\cal E}_0$ (Fig.3). As a
result, existence of the phase transition
becomes to be proved
on the formal level (Sec.5).
This conclusion should not be considered as something incredible:
the model suggested by Aubry and Andre \cite{400} gives an
example of 1D system, exhibiting the Anderson transition with
the usual scaling properties \cite{401}.

\begin{figure*}
\centerline{\includegraphics[width=6.5 in]{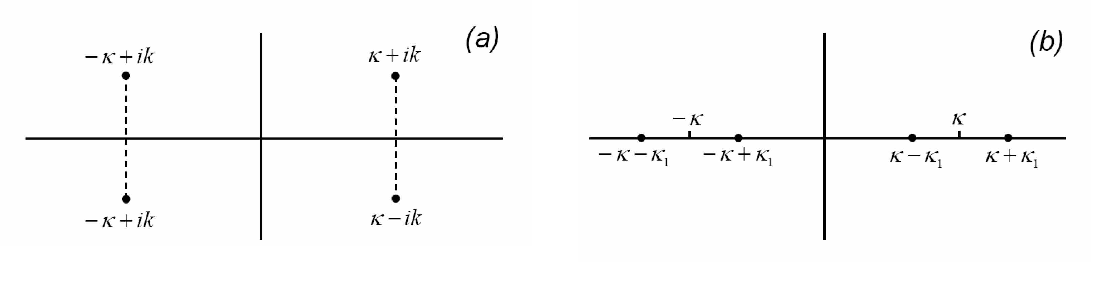}}
\caption{
{\small  The sense of the phase transition at the point
${\cal E}_0$ consists in the change
$i\langle k\rangle \to \langle \kappa_1\rangle$, so instead
of two pairs complex-conjugated exponents for
${\cal E}>{\cal E}_0$ $(a)$, one has four real exponents for
${\cal E}<{\cal E}_0$ $(b)$.  } }
\label{fig3}
\end{figure*}
%

\begin{center}
{\bf 2. Problems with the Gaussian distribution for $z_1$
and $z_2$.}
\end{center}

Let verify the hypothesis on the Gaussian distribution
for
$z_1$ and $z_2$, not specifying the choice of
solutions $y_1$ and $y_2$. Defining $z_1$ and $z_2$
according (13), and their average values according (14),
we accept for them the mutual Gaussian distribution
$$
P(z_1,z_2)\sim\exp\left\{ -\frac{1}{n} \left[\, B_{11}\tilde
z_1^2+ 2 B_{12}\tilde z_1 \tilde z_2 +B_{22}\tilde z_2^2\,
\right]  \right\}\,,
\eqno(21)
$$
where
$$
\tilde z_1=z_1-v_1 n\,, \qquad \tilde z_2=z_2-v_2 n\,.
\eqno(22)
$$
Then it easy to derive, that the moments of $y_1$ and $y_2$
have an exponential behavior
$$
\left\langle y_1^{m_1} y_2^{m_2} \right\rangle=
\exp\left\{ \kappa_{m_1 m_2} n \right\}
\eqno(23)
$$
with the exponents
$$
 \kappa_{m_1 m_2} = m_1 v_1 +m_2 v_2 + \,
 \frac{\,A_{11} m_1^2+ 2 A_{12} m_1 m_2 +A_{22} m_2^2}{4} \,,
\eqno(24)
$$
where $|| A_{ij}||$ is the matrix inverse to $|| B_{ij}||$.
In the paper \cite{123} we have found the complete set of
exponents for the moments $\left\langle \rho^n \right\rangle$
in the deep of the forbidden band
$$
 \kappa_{n,k} = 2(n-k)\delta + \epsilon^2
\left[ 2 n^2 -n -6 nk +3k^2\right]\,,
$$
$$
\qquad
k=0,1,\ldots,2n \,.
\eqno(25)
$$
Since the moments $\left\langle \rho^n \right\rangle$
are determined by averages (23) with $m_1+m_2=2n$,  we
can  set
$$
m_1=k\,, \qquad m_2=2n-k
\eqno(26)
$$
and obtain the result
$$
 \kappa_{m_1 m_2} = (m_1-m_2) \delta -
\frac{1}{2}\epsilon^2 (m_1+m_2)+
$$
$$
  + \frac{1}{2}\epsilon^2 \left[\,  m_1^2 -4 m_1 m_2  + m_2^2\,
  \right] \,,
\eqno(27)
$$
which is described by Eq.24 with parameters
$$
v_1=\delta-\epsilon^2/2\,, \quad v_2=-\delta-\epsilon^2/2\,,
\quad
$$
$$
A_{11}=2\epsilon^2\,, \quad A_{12}=-4\epsilon^2\,, \quad
A_{22}=2\epsilon^2\,
\eqno(28)
$$
and leads to the distribution
$$
P(z_1,z_2)\sim\exp\left\{ \frac{\tilde z_1^2+
4\tilde z_1 \tilde z_2 +\tilde z_2^2}{6\epsilon^2 n} \right\}\,.
\eqno(29)
$$
The determinant of the quadratic form in the
exponential of (29) is negative, and calculation of averages
requires rotation of the integration contour into the
complex plane: a physical sense of it is clarified in Sec.4.
Integration of (29) over $z_1$ or $z_2$ leads to results
$$
P(z_1)\sim\exp\left\{ -\frac{ (z_1-v_1 n)^2}{2\epsilon^2 n}
\right\}\,,
$$
$$
P(z_2)\sim\exp\left\{ -\frac{ (z_2-v_2 n)^2}{2\epsilon^2 n}
\right\}\,,
\eqno(30)
$$
the first of which reproduces the correct distribution for
the growing Lyapunov exponent \cite{13}. The result for the
decreasing exponent looks rather strange, since it
violates the condition $v_1=-v_2$, following from the
discussion in Sec.1. This point is clarified in Sec.3.

Another strange point is related with transition to the
allowed band. Since the equations for parameters
$\kappa_{m_1 m_2}$ are regular in energy \cite{13}, such
transition in formulas (25), (27) is produced by the changes
 $\delta \to i\delta$, $\epsilon \to i\epsilon $.
The same changes in (28) and (30) lead to correct values
$v_1=i\delta+\epsilon^2/2$,  $v_2=-i\delta+\epsilon^2/2$
in the allowed band, but do not reproduce the proper value
for $D$ in Eq.12. The latter is explained by the fact that
relationship of moments with the Lyapunov exponents is different
in the allowed and forbidden bands (Secs.4,5).

The Gaussian distribution (21) is valid not
for all energies.  It is easy to verify
(Sec.4), that for the energy ${\cal E}=0$, corresponding to the
initial band edge, the sets of the exponents for the second
$$
2\eta_0\,,\qquad \eta_0(-1+i\sqrt{3})\,,\qquad
\eta_0(-1-i\sqrt{3})
\eqno(31)
$$
and fourth moments
$$
2\eta_1\,,\quad 0\,, \quad 0\,, \quad
\eta_1(-1+i\sqrt{3})\,,\quad \eta_1(-1-i\sqrt{3}) \,,
\eqno(32)
$$
where \mbox{$\eta_0=(\epsilon^2 \delta^2)^{1/3}$},
\mbox{$\eta_1=(21\epsilon^2 \delta^2)^{1/3}$}, are not described
by  Eq.24.
The reasons for deviations from the
Gaussian form are discussed in Sec.6.

\begin{center}
{\bf 3. The structure of solution in the coordinate
space.} \end{center}

From definition (11) for $z$, one has equalities
$$
\sqrt{\rho+1}={\rm ch} (z/2)\,, \qquad
\sqrt{\rho}={\rm sh} (z/2)\,,
\eqno(33)
$$
relating the transfer matrix $T$ with the growing
Lyapunov exponent. If a wave falls from the left
on the system of point scatterers (Fig.1), then it is partially
reflected at each of them. As a result, at each point of the
system one has a superposition of waves propagating in opposite
directions. Let choose the point $x=l$ in the interval between
$n$th and $(n\!+\!1)$th scatterers.  Solution of the
Schr${\rm\ddot o}$dinger
equation on the left of the system ($x<0$)
$$
y(x)=Ae^{ikx}+Be^{-ikx}
\eqno(34)
$$
is related with a solution in the vicinity of $x=l$
by the transfer matrix for a system of $n$ scatterers
$$
y(x)=Ce^{ikx}+De^{-ikx}=
\eqno(35)
$$
$$
=\left[ A e^{i\varphi}{\rm ch} (z/2) \! + \!
B e^{i\theta}{\rm sh} (z/2)  \right] e^{ikx} +
$$
$$
+\left[ A e^{-i\theta}{\rm sh} (z/2) \! + \!
B e^{-i\varphi}{\rm ch} (z/2)  \right] e^{-ikx} \,.
$$
Using the combined phases (20), we can rewrite (35) in the form
$$
\!\!\!\!\!\!\!\!\!\!y(x)
\! =\!\!A\! \left[  e^{z/2-i\psi/2}\! \cos{(kx\!+\!\!\chi/2)}\! + \!
i e^{-z/2-i\psi/2} \sin{(kx\!+\!\!\chi/2)}\! \right]
$$
$$
\!\!\!\!+B \left[  e^{z/2+i\psi/2} \!\cos{(kx\!+\!\!\chi/2)} -
i e^{-z/2+i\psi/2} \sin{(kx\!+\!\!\chi/2)} \right] \,.
\eqno(36)
$$
Setting \mbox{$A=1$, $B=0$} or \mbox{$A=0$, $B=1$}, we have
two linearly independent solutions. In the transfer matrix
description, the part of the system in the interval $(0,l)$
is replaced by the effective point scatterer; the coordinate
$x$ for the arising effective system accepts a fixed value
\mbox{($x=0$)}, while the change of  $l$ is taken into account
by the change of the number of scatterers $n$. For a large
concentration of weak scatterers (Sec.7) the quantity  $n a_0$
becomes the continuous variable, replacing the coordinate $x$.
The choice of the $x$ origin is arbitrary, and instead of $x=0$
one can set \mbox{$x=x_0$} with \mbox{$x_0\sim 1$}. Since $x_0$
can be chosen differently for two linearly-independent
solutions, the latter can be written in the form
$$
y_1=   e^{z/2-i\psi/2} \cos{(\chi_1/2)} +
i e^{-z/2-i\psi/2} \sin{(\chi_1/2)} \,,
$$
$$
\eqno(37)
$$
$$
y_2=  e^{z/2+i\psi/2} \cos{(\chi_2/2)} -
i e^{-z/2+i\psi/2} \sin{(\chi_2/2)}  \,,
$$
where $\chi_1$ and $\chi_2$ differ from $\chi$ by the quantity
of the order of unity. According to \cite{125}, the phase
$\chi$ has no systematic growth, and its fluctuations remain
restricted.
The average values of $z$ and $\psi$
increase linearly with the number of scatterers
$n$ (see below), and $n$ plays the role of the coordinate  $x$ in
Eq.4. Then a linear combination of two solutions (37) corresponds
to a superposition of four exponents in Eq.19.\,\footnote{\,As
was already indicated and will be clear from the following,
behavior of parameters $\kappa$ and $k$ is essentially different
at large and small scales.  At small scales they are strongly
fluctuating (so that solution (19) satisfies the
Schr${\rm\ddot o}$dinger
equation at all scales), while at large scales tend to the
constant values. This difference is a reason of paradoxes
indicated in Sec.1, which consequently have the objective
character.  The mentioned constant values are essential for
justification of the phase transition (Sec.5).}
One can see, that
coefficients $C_i$ in Eq.19 cannot be chosen independently;
as a result, a partial solution has not a
form of a single exponent, but always contains their
superposition. Approximate reducing to a single
exponent is possible, if fluctuations of the phase $\chi$
are neglected: then accepting for $\chi_1$ and $\chi_2$
values  $0$ or $\pi$, one can eliminate one of the exponents
in $y_1$ and $y_2$, and provide the constant value
of the Wronskian (3). In the general case, the fixed value of the
Wronskian is provided by the fact, that variations of $z$ and
$\psi$ are separated in space: in the intervals between
scatterers the quantity $z$ is constant, while
variations of $\psi$ correspond to superposition of
solutions  $e^{\pm ikx}$ or $e^{\pm \kappa x}$, depending
on the position of the band edge shifted due to fluctuations;
contrary, $z$ changes in the vicinity of scatterers,
while $\psi$  remains practically constant. If a random potential
changes smoothly and cannot be represented as a succession of
scatterers, then essential changes of $z$ occur near
the turning points, while in absence of the latter they appear to
be very slow and remain within uncertainty of exponents related
with variation of $\psi$. Formulas (37) allow to understand the
mechanism of appearance of localized states: e.g. the solution
$y_1$ becomes restricted in the whole space, if $\chi_1\to\pi$
for $x\to\infty$ and $\chi_1\to 0$ for $x\to-\infty$.

Let return to the statement on the linear growth of the
average values of $z$ and $\psi$. This statement follows from the
recurrence relations describing the change of parameters
of the transfer matrix $T$, when a number of scatterers is
increased by unity
\cite{126}\,\footnote{\,The first relation follows from
Eq.25 of the paper  \cite{126} after substitution $\rho=\exp{z}$,
the second one arises from Eq.29 of the same paper for $R=2$. In
both cases we accept the value $-\pi/2$ for the parameter
$\gamma$, which correspond to abrupt boundaries between the
system and the ideal leads. Parameters $\alpha$,
$\beta$, $\gamma$, $\Delta$ are introduced in Eq.18 of the
indicated paper.}
$$
z_{n+1}=z_n +2\Delta \cos(\psi_n\!-\!\beta)
-2\epsilon_n\sin \psi_n+
$$
$$
+2\epsilon_n^2 (\cos^2 \psi_n-\cos \psi_n)\,,
\eqno(38)
$$
$$
 \psi_{n+1}=\psi_n -2\alpha-2\Delta \sin(\psi_n\!-\!\beta)
+2\epsilon_n(1-\cos \psi_n) + $$ $$ +2\epsilon_n^2
\sin\psi_n(1-\cos \psi_n)\,.
$$
Here  $\epsilon_n$ are statistically independent of $\psi_n$
and proportional to the  energies of sites in
the 1D Anderson model; they have zero means and equal
variances, $\left\langle \epsilon_n \right\rangle=0$,
$\left\langle \epsilon_n^2 \right\rangle=\epsilon^2$.  It is easy
to see that  $z_n$ and $\psi_n$ are represented by sums of $n$
random quantities with approximately equal distributions: the
mutual Gaussian distribution is rather likely for them, while
their means and variances grow linearly in $n$. In particular,
for the mean and variance of $z_n$ one has
$$
\left\langle z_n \right\rangle =
n \left[ 2\Delta \left\langle\cos(\psi\!-\!\beta)
\right\rangle
+2\epsilon^2 \left\langle\cos^2 \psi-\cos \psi\right\rangle
\right]\equiv vn  \,,
$$
$$
\left\langle \left( \delta z_n \right)^2
\right\rangle
= 4\epsilon^2 \left\langle\sin^2 \psi
\right\rangle n \equiv 2Dn
\eqno(39)
$$
in correspondence with the parameters $v$ and $D$ of
the log-normal distribution for $\rho$ \cite{126}. The
distribution of the phase $\psi$ is stationary \cite{126},
if it reduces to the interval $(0,2\pi)$.
If, however, $\psi$ is defined
by continuity, then its mean grows proportionally to $n$:
$$\!\!\!\!
\left\langle \psi_n \right\rangle \!=
n \left[\!-2\alpha\!-\! 2\Delta \left\langle\sin(\psi\!-\!\beta)
\right\rangle\! +\!2\epsilon^2 \left\langle\sin \psi(1-\cos
\psi)\right\rangle \right],
\eqno(40)
$$
and estimation of the right-hand part in the random phase
approximation gives $\left\langle \psi_n \right\rangle=-2\alpha
n$, and in the deep of the allowed band it reduces to
$\left\langle \psi_n \right\rangle=2n\delta$ or $\left\langle
\psi \right\rangle=2 kx$. Generally, the same relation can be
accepted by definition with the momentum $k$, renormalized due to
disorder.

\begin{center}
{\bf 4. General analysis for the Gaussian distribution.}
\end{center}

In general, the quantities $z_1$ and $z_2$ are complex-valued.
In order to deal with real distributions, let accept the
most general form for $y_1$ and $y_2$
$$
y_1={\rm e}^{iS_1+S_2+iS_3+S_4}\,, \quad
y_2={\rm e}^{-iS_1-S_2+iS_3+S_4}\,,
\eqno(41)
$$
where $S_i$ are sums of $n$ random quantities with
average values
$$
\left\langle  S_i   \right\rangle = v_i n\,,
\eqno(42)
$$
and expected to obey the mutual Gaussian distribution
$$
P\{S_i\} \sim \exp \left\{ -\frac{1}{n} \sum_{ij} B_{ij}
\left( S_i-v_i n \right) \left( S_j-v_j n \right) \right\}
\,.
\eqno(43)
$$
Then for the moments
$\left\langle y_1^{m_1} y_2^{m_2} \right\rangle$ one can
obtain the exponential behavior (23) with the exponents
(see Appendix 1)
$$
\kappa_{m_1 m_2} = (m_1\!-\!m_2) (i v_1+v_2) +
(m_1\!+\!m_2) (i v_3+v_4) +
$$
$$
+\frac{(m_1\!-\!m_2)^2}{4} \left( -A_{11}+2iA_{12}+A_{22}\right)+
$$
$$
+\frac{m_1^2\!-\!m_2^2}{2} \left( -A_{13}+iA_{14}+iA_{23}+A_{24}\right)
+
$$
$$
+\frac{(m_1\!+\!m_2)^2}{4} \left( -A_{33}+2iA_{34}+A_{44}\right)\,,
\eqno(44)
$$
where the matrix $|| A_{ij} ||$ is inverse to $|| B_{ij} ||$.

\vspace{2mm}

In formulas (37) for $y_1$ and $y_2$ the first terms
increase with a number of scatterers, and it is natural to expect
that namely these terms are responsible for evolution of
moments at large $n$. In what follows, we demonstrate that it so
indeed.

\vspace{4mm}
{\it Allowed band.}
\vspace{2mm}

\begin{figure}
\centerline{\includegraphics[width=2.6 in]{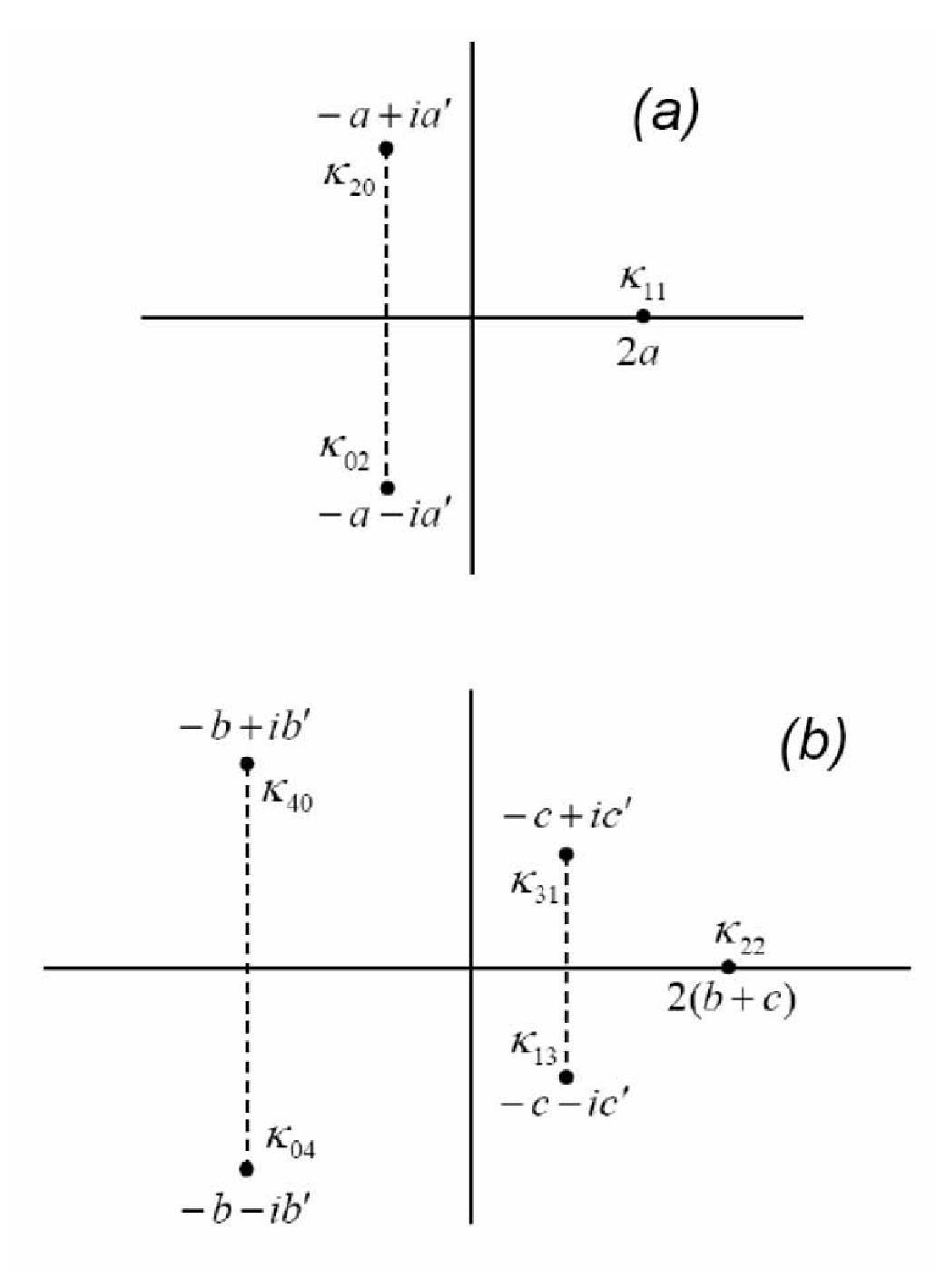}}
\caption{ \small  Configuration of exponents $\kappa_{m_1 m_2}$
for the second and fourth moments in the deep of the allowed
band. A sum of the exponents is equal to zero due to the
generalized Viete theorem (see Appendix2). }
\label{fig4} \end{figure}

Retaining the first terms in Eqs.37 and compare with (41),
we can set
$$
S_1=\psi/2\,, \quad S_2=0\,, \quad  S_3=0\,, \quad S_4=z/2\,,
\eqno(45)
$$
and all parameters with indices 2 and 3 vanish in Eq.44:
$$
\kappa_{m_1 m_2} = (m_1\!-\!m_2) i v_1 + (m_1\!+\!m_2) v_4
-\frac{(m_1\!-\!m_2)^2}{4}\,  A_{11}+
$$
$$+
\frac{m_1^2\!-\!m_2^2}{2}\, i  A_{14}
+\frac{(m_1\!+\!m_2)^2}{4} \, A_{44} \,.
\eqno(46)
$$
Configurations of exponents $\kappa_{m_1 m_2}$ for the second and
fourth moments in the deep of the allowed band are shown in Fig.4;
a sum of the exponents is equal to zero due to the generalized
Viete theorem (see Appendix 2). Let introduce the symmetric and
antisymmetric combinations in respect to permutation of $m_1$
and $m_2$,
$$
\!\!\tilde\kappa_{m_1 m_2}=
\frac{\kappa_{m_1 m_2}\!\!-\!\kappa_{m_1 m_2}}{2\,i}\,,
\quad
\tilde{\tilde\kappa}_{m_1 m_2}=
\frac{\kappa_{m_1 m_2}\!\!+\!\kappa_{m_1 m_2}}{2}\,,
\eqno(47)
$$
which are determined by different sets of  coefficients.
Using parametrization of exponents, introducing by Fig.4, we have
the equations
$$
\tilde{\tilde\kappa}_{1 1} = 2 v_4 + A_{44}= 2a \,,
$$
$$
\tilde{\tilde\kappa}_{2 0} = 2 v_4 -   A_{11}+  A_{44}
=-a \,,
$$
$$
\tilde{\tilde\kappa}_{2 2} = 4 v_4 + 4  A_{44} = 2b+2c \,,
\eqno(48)
$$
$$
\tilde{\tilde\kappa}_{4 0}= 4 v_4 - 4 A_{11}+ 4 A_{44} = -b \,,
$$
$$
\tilde{\tilde\kappa}_{3 1}= 4 v_4 -   A_{11}+ 4 A_{44} = -c \,,
$$
$$
{\tilde\kappa}_{2 0}= 2  v_1 + 2  A_{14} =a' \,,
$$
$$
{\tilde\kappa}_{4 0}= 4  v_1 + 8   A_{14} =  b' \,,
$$
$$
{\tilde\kappa}_{3 1} = 2  v_1 + 4   A_{14} =c'\,,
$$
which lead to results for parameters
$$
v_4=-c/6\,, \quad A_{11} =-c\,,
\quad A_{44} =-c/3\,,
$$
$$
v_1=a'-c'/2\,, \quad  A_{14} =(c'-a')/2
\eqno(49)
$$
and conditions of self-consistency
$$
 3a=-c\,,\quad  b=-2c\,, \quad b'=2c'\,.
\eqno(50)
$$
In the extremely metallic regime we have \cite{13}
$$
a=\epsilon^2\,, \quad b=6\epsilon^2\,, \quad c=-3\epsilon^2\,,
$$
$$
 a'=2\delta\,, \quad b'=4\delta\,, \quad c'=2\delta\,,
\eqno(51)
$$
so that self-consistency conditions are fulfilled, and the
hypothesis on the Gaussian distribution is confirmed.
For parameters of Eq.46 we obtain
$$
v_1=\delta\,, \quad v_4=\epsilon^2/2\,,
$$
$$
A_{11} =3\epsilon^2\,, \quad  A_{14} =0\,,
\quad  A_{44} =\epsilon^2\,,
\eqno(52)
$$
so that
$$
 \kappa_{m_1 m_2} = m_1(i \delta +\epsilon^2/2)
 +m_2(-i \delta +\epsilon^2/2)-
$$
$$
  -\frac{m_1^2 -4 m_1 m_2  + m_2^2}{2}\, \epsilon^2   \,.
\eqno(53)
$$
After replacement $i\delta\to \delta$, $i\epsilon\to \epsilon$
it reproduces the result  (27), derived in \cite{123} on the
base of the  "diagonal" approximation (see Appendix 2).

Since the matrix $||A_{ij}||$ is diagonal, we have
$B_{ii}=1/A_{ii}$, and distribution (43) accepts the form
$$
P\{S_i\} \sim \exp \left\{ -\frac{(S_1-n\delta)^2}{3\epsilon^2 n}
-\frac{(S_4-n\epsilon^2/2)^2}{\epsilon^2 n}
 \right\} \,.
\eqno(54)
$$
Integrating over $S_1$ and substituting $S_4=z/2$,  we come
to the distribution of $z$
$$
P\{z\} \sim \exp \left\{
-\frac{(z-n\epsilon^2)^2}{4\epsilon^2 n}
 \right\} \,,
\eqno(55)
$$
corresponding to the log-normal distribution for $\rho$
with parameters $v=\epsilon^2$, $D=\epsilon^2$,
valid in the deep of the allowed band \cite{13}. The average
value  $\left\langle S_1 \right\rangle=n\delta$ is in agreement
with the results for $\left\langle \psi/2 \right\rangle$,
given in the previous section. For fluctuations of $\psi_n$ we
can derive from (38)
$$
\delta\psi_{n}=\sum\limits_{k=1}^{n} 2\epsilon_k(1-\cos \psi_k)
+O(\Delta, \epsilon^2)\,,
\eqno(56)
$$
and estimation of the average
$$
\left\langle \left(\delta\psi_{n}\right)^2  \right\rangle
=4\epsilon^2 n
\left\langle 1-2\cos \psi + \cos^2 \psi  \right\rangle
\eqno(57)
$$
in the random phase approximation  gives $\left\langle
\left(\delta\psi_{n}\right)^2 \right\rangle= 6\epsilon^2 n$, in
correspondence with the variance of $S_1$ following from (54).
The distribution (54) justifies the self-averaging property for
the imaginary part of the Lyapunov exponents.

\vspace{4mm}

{\it Forbidden band.}
\vspace{2mm}

According to  \cite{125,126},  at certain energy  ${\cal E}_0$
the phase $\psi$  acquires an imaginary part,
$\psi=\psi'+i\psi''$, and we have from Eq.37, retaining the first
terms
$$
y_1={\rm e}^{z/2-i\psi'/2+\psi''/2}\,, \quad
y_2={\rm e}^{z/2+i\psi'/2-\psi''/2} \,.
\eqno(58)
$$
In the case $\psi''=0$, solutions  $y_1$ and $y_2$ were
complex-conjugated, and taking their sum and difference
one can present the solutions in the real form.
In the presence of $\psi''$, the moduluses $|y_1|$ and $|y_2|$
grow with different rates\,\footnote{\,We have in mind the
"essential" imaginary part of $\psi$, whose
average value grows proportionally to $n$.},
and the real part $\psi'$ should disappear, since in the
opposite case one cannot obtain real solutions.  Comparing with
(41), we have
 $$
 S_1=0\,, \quad S_2=\psi''/2\,, \quad  S_3=0\,, \quad
S_4=z/2\,,
\eqno(59)
$$
and all parameters with indices 1 and 3  vanish in Eq.44,
i.e.
$$
\kappa_{m_1 m_2} = (m_1\!-\!m_2) v_2 + (m_1\!+\!m_2) v_4+
\frac{(m_1\!-\!m_2)^2}{4}\, A_{22}+
$$
$$
+\frac{m_1^2\!-\!m_2^2}{2}\, A_{24}
+\frac{(m_1\!+\!m_2)^2}{4} \, A_{44} \,.
\eqno(60)
$$
The exponents $\kappa_{m_1 m_2}$ in the forbidden
band differ from those in the allowed band by the changes
$i\delta\to \delta$, $i\epsilon\to \epsilon$, and proceeding
analogously, one has for parameters
$$
v_2=\delta\,, \quad v_4=-\epsilon^2/2\,,
$$
$$
A_{22} =3\epsilon^2\,, \quad  A_{24} =0\,,
\quad  A_{44} =-\epsilon^2\,,
\eqno(61)
$$
and their substitution to (60) returns to (27).

The matrix $||A_{ij}||$ is diagonal, so $B_{ii}=1/A_{ii}$,
and the distribution (43) accepts the form, analogous to (54)
$$
P\{S_i\} \sim \exp \left\{ -\frac{(S_2-n\delta)^2}{3\epsilon^2 n}
+\frac{(S_4+n\epsilon^2/2)^2}{\epsilon^2 n}
 \right\} \,.
\eqno(62)
$$
However, such representation is inconvenient, since the
growing Lyapunov exponent is determined not by the quantity
$S_4$, but the quantity $S_4+S_2$. Setting
$$
z_1=S_2+S_4\,,\qquad  z_2=-S_2+S_4\,,
\eqno(63)
$$
we return to the result (29) with parameters $v_i$ from (28),
reproducing the correct distribution (31) for the growing
Lyapunov exponent. According to (62), the quantity  $S_4$
possesses a real mean, but a pure imaginary fluctuation: it
leads to the negative determinant of the quadratic form in (29)
and complex fluctuations of $z_1$ and $z_2$.

\vspace{4mm}

{\it The boundary of the initial band.}
\vspace{2mm}

According to (31),(32), configurations of exponents $\kappa_{m_1
m_2}$ for the energy ${\cal E}=0$, corresponding to the initial
band edge, are given by Fig.4 with parameters
$$
a=\eta_0\,, \quad b=\eta_1\,, \quad c=0\,,
$$
$$
 a'=\eta_0\sqrt{3}\,, \quad b'=\eta_1\sqrt{3}\,, \quad
c'=0\,.
$$
It is easy to see that conditions of self-consistency (50)
are not fulfilled, and the hypothesis on the Gaussian
distribution is not confirmed.
\vspace{4mm}

Analysis of the present and previous sections allows to
identify the actual pair of exponents, responsible for
evolution of moments, and accept them as $y_1$ and $y_2$.
This is a crucial step in establishing the mutual distribution
$P(y_1,y_2)$. Initially, it looks evident that in the
capacity of $y_1$ and $y_2$ one should take the growing and
decreasing Lyapunov exponents with opposite exponentials,
while the main  problem consists in revelation of the decreasing
exponent in the background of the growing solution and its formal
definition. In fact, a situation is different: if the
Schr${\rm\ddot o}$dinger equation is integrated from left to
right, then the first  two terms in Eq.19 are dominated, while
for the opposite integration two last terms are actual.  It
explains violation of equality $v_1=-v_2$ in the Gaussian
distribution for $z_1$ and $z_2$ (Sec.2). The formal definition
of actual exponents is given by Eq.37.

\begin{center}
{\bf 5. Consequences for the phase transition in the
distribution $P(\psi)$ }
\end{center}

The above analysis provides new information on the phase
transition in the distribution of $\psi$, predicted
in \cite{125,126}. This transition consists in appearance of
the imaginary part for the phase $\psi$, related with inevitable
transformation of the true transfer matrix $T$,
describing a probe scatterer in the allowed band, to the
pseudo transfer matrix $t$, describing a scatterer in the
forbidden band \cite{13}. The difference between two types of
matrices can be made arbitrary large, if their separation in
energy is increased, and this difference cannot be overcome by
addition of a weak random potential. As a result, the border-line
between the true and pseudo transfer matrices can be only shifted
but not eliminated.

According to Sec.4, appearance of the imaginary part of $\psi$
is accompanied by disappearance of its real part $\psi'$.
Due to relation  $\left\langle \psi \right\rangle=2kx$,
it is analogous to a situation in the ideal crystal, where
transition from the allowed to forbidden band reduces to the
change $k\to i\kappa$. In the disordered system, the analogous
change is valid for average values,
$\left\langle k \right\rangle\to i\left\langle \kappa_1
\right\rangle$, and instead of two pairs of the
complex-conjugated exponents in Eq.19 we have four real exponents
(Fig.3).\,\footnote{\,In another context, the difference
between the pure  real and complex-valued wavenumber arouse in
the approach of papers \cite{402}.}
Self-averaging of all four Lyapunov exponents follows
from distributions (54), (62). Realization of two different
configurations is established by the above analysis in the deep
of the allowed and forbidden bands: it proves the existence of
singularity in energy on the formal
level\,\footnote{\,Existence of a singularity at
the point  ${\cal E}_0$ is established in \cite{124}--\cite{126}
on the base of the physical reasoning with the use of
numerical analysis, so that the formal arguments were practically
absent.}.  Therefore, a difference between the allowed and
forbidden band survives in disordered systems, though a
singularity in the density of states is smeared out.
It resembles the famous argumentation by Mott \cite{2}, that
the role of the allowed band edge comes to the mobility
edge; the latter is absent in the 1D  case, but a 'trace' of it
still remains.

A difference of two situations is manifested in the
behavior of the Cauchy solution for given
initial conditions on one of the ends of the system. For
${\cal E}>{\cal E}_0$ the mean  $\langle k\rangle$ is
finite, and the change of a sign of such solution occurs
regularly on the scale $1/\langle k\rangle$. A situation
for ${\cal E}<{\cal E}_0$ is clearly understood for
energies in the deep of the forbidden band: then one has a
quickly growing Lyapunov exponent, with essential fluctuations
around it, related with disorder. It remains always
possible, that one of fluctuations extends
till zero, providing a fluctuational change of the sign.
However,  with the growth of the exponent such
events occur more infrequently
and no characteristic scale can be
related with them; hence, one cannot reveal any
finite value of $\langle k\rangle$.

In other words, in the ideal system the allowed
and forbidden band differ in two aspects: (i) the Cauchy
solution is restricted in the former case and growing in
the latter; (ii) the solution is oscillating in the first
case, and changes monotonically in the second one. When a
disorder is added to the system, the
difference disappears in relation of the former aspect
(the Cauchy solution grows in both cases), but retains in respect
of the latter.  Correspondingly, resistance and density of states
become regular in  energy, while the phase analysis allows to
register the transition. The above arguments
essentially simplify registration of the transition in optical
systems \cite{124,125}:  statistical analysis of $\psi$ becomes
unnecessary, and it is sufficient to trace a change of a sign of
the field in the wave, while moving along the coordinate.

It should be clear, that the average
$\langle\psi\rangle$ tends to zero at approaching the point
${\cal E}_0$, and solutions $y_1$ and $y_2$ become coinciding. It
is analogous to a situation for the usual second order transitions,
when two quadratic minima in the free energy
approach each other and transform to the minimum of the fourth
order.  Correspondingly, the Gaussian fluctuations around remote
quadratic minima become non-Gaussian at their approaching.
According to Sec.7, deviations from the Gaussian form are
indeed related with the $\psi$ distribution, while
fluctuations of $z$ remain always Gaussian.

The appearance of the imaginary part
of $\psi$ changes correspondence between the moments of the
Cauchy solution and the Lyapunov exponents: according to (58), for
$\psi''>0$  the solution  $y_1$ grows faster than $y_2$, and the
growing Lyapunov exponent is determined by the moments
$\left\langle y_1^{2m}y_2^0\right\rangle$ for \mbox{${\cal
E}<{\cal E}_0$}, while for \mbox{${\cal E}>{\cal E}_0$} it was
related with  the moments  $\left\langle
y_1^{m}y_2^m\right\rangle$. Such change of the regime is
confirmed by the above results for the allowed and forbidden
bands.

Next, appearance of the imaginary part of $\psi$ changes
definition of the Lyapunov exponents. If phases $\varphi$
and $\theta$ in the transfer matrix (7) become complex-valued,
$$
\varphi=\varphi'+i\varphi''\,,\qquad
\theta=\theta'+i\theta''\,,
\eqno(64)
$$
then it transforms to the pseudo transfer matrix
$$
 t= \left ( \begin{array}{cc}
\!\!\sqrt{\bar\rho\!+\!1}\, e^{i\varphi'-\varphi''}\!\! &
\!\!\sqrt{\bar\rho} \,e^{i\theta'-\theta''}\!\!
\\ \!\!\sqrt{\bar\rho}\, e^{-i\theta'+\theta''}\!\!
&\!\! \sqrt{\bar\rho\!+\!1}\,
e^{-i\varphi'+\varphi''}\!\! \end{array} \right)\,,
\eqno(65)
$$
where the parameter $\bar\rho$ can be different from
the Landauer resistance $\rho$.  Composing the Hermitian
matrix
$$
t t^+=
$$
$$\!\!\!\!
\left ( \!\begin{array}{cc}
\!\!{\!\!\!\!\!\!\!\!\!\!\!\!\!\!(\bar\rho\!+\!1)}\, e^{-2\varphi''}\!\!
\!\!+\!\bar\rho\, e^{-2\theta''} & \!\!\!\!\!\!\!\!
\!\!\!\!\!\!\!\!\!2\sqrt{\bar\rho(\bar\rho\!+\!1)}
\,e^{i\varphi'+i\theta'}\!\!{\rm ch}(\theta''\!\!-\!\varphi'')\!\!
\\
\!\!2\sqrt{\bar\rho(\bar\rho\!+\!1)}
\,e^{-i\varphi'-i\theta'}\!\!{\rm ch}(\theta''\!\!-\!\varphi'')\!\!
&{(\bar\rho\!+\!1)}\, e^{2\varphi''}\!\!\!\!
+\!\bar\rho\, e^{2\theta''} \!\! \end{array} \!\right)\!,
\eqno(66)
$$
we have the following equation for its eigenvalues
$\lambda_{1,2}=\exp(\pm z)$
$$
{\rm ch}z=(\bar\rho\!+\!1)\,{\rm ch}2\varphi''+
\bar\rho\,{\rm ch} 2\theta''\qquad (z>0) \,,
\eqno(67)
$$
which for large  $\bar\rho$ accepts the form
$$
{\rm ch}z=2\bar\rho \,{\rm ch}\psi''{\rm ch} \chi''\,.
\eqno(68)
$$
The imaginary part of the phase $\chi$ is forbidden by
flux conservation \cite{124,125}, so $\chi''=0$. The average
value of $\psi$ grows linearly with the number of
scatterers $n$, and setting
$$
\bar \rho = {\rm e}^{\bar z} \qquad  (\bar z>0)
\eqno(69)
$$
we have the following relation for large $n$
$$
z=\bar z + |\psi''| \,,
\eqno(70)
$$
which
manifests
redefinition of the Lyapunov exponents.

However, relationship of the growing Lyapunov exponent with
the Landauer resistance $\rho$ remains unchanged. Indeed,
for $\chi''=0$ the elements of the matrix $t$ has
an order of growth $\exp \left(\bar z/2 \pm \psi''/2 \right)$,
and dependently on the sign of $\psi''$ dominates either
right ($\psi''>0$), or the left ($\psi''<0$) column. Transition
to the true transfer matrix $T$ (determining $\rho$) is
given by relation
$$
T=T_l\, t\, T_r \,,
\eqno(71)
$$
where $T_l$ and $T_r$ are the constant edge matrices \cite{13}.
The elements of $T$ are determined by the linear
combinations of the $t$ elements, which are dominated by the most
quickly growing terms. Since the moduluses of the $T$
elements grow as $\sqrt{\rho}$, then
$\sqrt{\rho}\sim \exp \left(\bar z/2 + |\psi''|/2 \right)$,
and the required relation $\rho\sim \exp{z}$ is established. This
conclusion agrees with the fact that the matrix $TT^+$ is
directly related with the resistance of the system, though with
somewhat different its definition \cite{200}.

The log-normal distribution for $\rho$ follows from the
evolution equation for $P(\rho)$ \cite{13}, and its parameters
$v$ and $D$ are regular in the energy (Fig.2),
providing regularity of the distribution for $z$, while
$\bar z$ and $\psi''$ have square-root singularities
\cite{124,125}. In fact, $\left\langle\psi''\right\rangle$
is an order parameter for this transition.

There remain some difficulties in the presented picture.
According to \cite{126}, the point  ${\cal E}_0$ is
situated inside the initial allowed band (Fig.2). If the
real part of $\psi$ disappears at the point ${\cal E}_0$,
then solutions $y_1$ and $y_2$ become real for
${\cal E}<{\cal E}_0$. However, at the energy ${\cal E}=0$,
corresponding to the initial band edge, some of the exponents
$\kappa_{m_1 m_2}$ still contain the imaginary part
(see Eq.32), and such situation retain for some negative energies.
This contradiction is resolved by the fact, that reality of $y_1$
and $y_2$ does not mean vanishing of $S_1$ and $S_3$ in Eq.41.
Solutions $y_1$ and $y_2$ can
change a sign, which leads to appearance of contributions
$\pm i\pi$ in exponentials of (41). If two solutions change
their signs in different points, then both $S_1$ and
$S_3$ remain finite. The latter quantities are not essential in
the framework of the Gaussian distribution, since their
neglect leads to the correct result for $z_1$ (Sec.4);
beyond it they play an important role, allowing to escape
contradictions (Sec.8). The average values of
$S_1$ and $S_3$ vanish, and their fluctuations grow as
$n^{1/4}$ instead of $n^{1/2}$ for the Gaussian distribution; the
latter reflects the fact that probability of the
fluctuational change of a sign reduces with growth of
the exponent.

\begin{center}
{\bf 6. Mechanism of deviations from the Gaussian
form.  }
\end{center}

As was already indicated, the Gaussian distribution does not
describe a situation for all energies. The reason for it is
easily clarified, if consideration is carried out in terms
of characteristic functions. In the case of one variable, the
characteristic function $F(t)$ is a Fourier transform of the
distribution function $P(x)$,
$$
F(t)=\int dx e^{ixt} P(x)\,,\quad
P(x)=\frac{1}{2\pi}\int dt e^{-ixt} F(t) \,,
\eqno(72)
$$
being the generating function of the moments
$\left\langle x^k \right\rangle$,
$$
F(t)=\left\langle e^{ixt} \right\rangle
=\sum_{k=0}^{\infty} \frac{(it)^k}{k!}\left\langle x^k
\right\rangle \,.
\eqno(73)
$$
The analogous relation for its logarithm is a definition of
the cumulants $\mu_k$:
$$
\ln F(t)=\sum_{k=1}^{\infty} \frac{(it)^k}{k!} \mu_k\,.
\eqno(74)
$$
Relationship of cumulants with moments is established by
taking the logarithm of the series (73); in particular,
$\mu_1=\langle x \rangle$, $\mu_2=\left\langle x^2
\right\rangle-\langle x \rangle^2$. Considering the average
of the exponent
$$
\left\langle e^{mx} \right\rangle =
\int e^{mx} P(x) dx   \,,
\eqno(75)
$$
it is easy to see, that it corresponds to the change
$it\to m$ in the definition of the characteristic function,
and the result is obtained trivially
$$
\left\langle e^{mx} \right\rangle = \exp\left\{
\sum_{k=1}^{\infty} \frac{\mu_k}{k!} m^k  \right\}\,.
\eqno(76)
$$
Let consider, what is happened in the course of summation of
random quantities. Validity of the central limit theorem is based
on the fact that characteristic functions are multiplied for
statistically independent  quantities. For a sum of $n$
equally distributed terms, the characteristic function
is obtained by the change $\mu_k\to \mu_k n$,
$$
F(t)= \exp\left\{\sum_{k=1}^{\infty} \frac{\mu_k n}{k!} (it)^k
\right\}=
$$
$$
=\exp\left(i \mu_1 n \, t - \frac{\mu_2 n}{2!} t^2
- i \frac{\mu_3 n}{3!}  t^3 +\ldots \right)\,.
\eqno(77)
$$
In calculation $P(x)$ by the inverse Fourier transform,
the integral over $t$ is restricted
by the second term in the exponential, and the main
contribution occurs
from the region $|t|\alt
n^{-1/2}$, where higher cumulants are small for large $n$, and can
be neglected to reveal the Gaussian distribution.  However, the
Gaussian form is valid only in the vicinity of the maximum
of the distribution, while its tails remain non-universal.

The latter has no significance in calculation  the
moments $\left\langle x^m \right\rangle$, but
becomes quite essential in the case of the exponential
averages.
For a sum of $n$ equally distributed quantities we have
a change $\mu_k\to \mu_k n$ in Eq.76,
$$
\left\langle e^{mx} \right\rangle = \exp\left\{
\sum_{k=1}^{\infty} \frac{\mu_k n}{k!} m^k  \right\}
\equiv \exp(\kappa_m n) \,,
\eqno(78)
$$
and the higher cumulants are essential for the exponent
$\kappa_m$ in the same degree, as they are essential for a single
term of the sum. We see that a central limit theorem is not
effective for the exponential averages.  Naturally, it is related
with the fact, that such averages are determined by the tails of
a distribution.

In the case of two variables, the characteristic function has a
structure
$$
F(t,t')=\exp \left\{ \sum_{k k'} \mu_{k k'} (it)^{k}
(it')^{k'}\right\} \,,\qquad k+k'\ge 1\,,
\eqno(79)
$$
and analogously for a greater number of variables; the
factorial coefficients are included in the definition of
cumulants.

\begin{center}
{\bf 7. Corrections to the diffusion equation and
influence of correlations.  }
\end{center}

Analysis of the previous section cast certain doubts on
applicability of the log-normal distribution for $\rho$ in
calculation the moments of $y_i$. This question is a matter of
principle, since the parameters of the log-normal distribution
(Fig.2) were established in \cite{13} by the analysis of
the second and fourth moments.

This point can be easily clarified. The first equation (38)
has a structure
$$
z_{n+1}=z_n - f\left( \psi_n \right)
\eqno(80)
$$
and the evolution equation for $P(z)$ can be derived
in the same manner, as for the distribution $P(\chi)$ in the
paper \cite{125}. Beyond the diffusion approximation,
this equation has a form
$$
\frac{\partial P}{\partial n} = D_1 P'_z +
D_2 P''_{zz} +  D_3 P'''_{zzz} +
\ldots,
$$
$$ D_k=\frac{1}{k!}\left\langle f^k(\psi)\right\rangle
\,,
\eqno(81)
$$
and can be easily solved in terms of the characteristic
function
$$
F(t)=\exp\left\{ n\sum_{k=1}^{\infty} \frac{1}{k!}
\left\langle f^k(\psi)\right\rangle (-it)^k \right\}\,.
\eqno(82)
$$
The cumulants of the distribution are determined by averages
of $f^k(\psi)$, and all of them are essential in the general
case.  However, beginning from  \cite{13}, in
all subsequent papers \cite{123,125,126} we consider the limit
$$
\delta\to 0\,,\quad \epsilon \to 0\,, \quad
\delta/\epsilon^2=const\,,
\eqno(83)
$$
when the terms of the
order of $\delta$, $\epsilon^2$, $\epsilon^4/\delta$,
$\epsilon^6/\delta^2$, $\ldots$ are retained, but contributions
$\epsilon^4$, $\epsilon^6$, $\ldots$ are neglected. According to
(38), $f(\psi)$ is a sum of terms containing
$\delta,\,\,\epsilon,\,\,\epsilon^2$. The terms of the order of
$\delta$ are present only in the first cumulant, the terms of
order $ \epsilon^2$ in the first and the second cumulant; the
third and the fourth cumulants begin with $\epsilon^4$, the
fifth and sixth ones begin with $\epsilon^6$, and so on. It is
clear that restriction by the first two cumulants is justified in
the limit (83).

By the same reasons, in the evolution equation for $P(\psi)$
one can retain only two derivatives over $\psi$.  However,
the obtained diffusion-type equation has coefficients
depending on $\psi$, and its solution is not Gaussian. In the
deep of the allowed and forbidden bands this equation can be
solved by iterations over $\epsilon^2/\delta$, so the main
contributions of the order of $\delta$ and $\epsilon^2$ should
be supplemented by terms $\epsilon^4/\delta$,
$\epsilon^6/\delta^2$,$\ldots$, which become essential
near the initial band edge.

The more detailed information can be obtained from the
analysis of correlations. According  to (38), the quantities
$z_n$ and $\psi_n$ are determined by sums of the form
$$
S=\sum_{k=1}^{n} \left( a_k \epsilon_k +b_k \epsilon_k^2 \right)
\,,
\eqno(84)
$$
where $a_k$ and $b_k$ are random quantities, independent of
$\epsilon_k$. For the first moment we have trivially
$$
\left\langle S \right\rangle =
\sum_{k} \left[ \overline{a_k} \left\langle\epsilon_k
\right\rangle +\overline{ b_k} \left\langle\epsilon_k^2
\right\rangle \right]=\bar b \epsilon^2 n \,.
\eqno(85)
$$
Calculating the second moment
$$
\left\langle S^2 \right\rangle =
\sum_{k}  \overline{a^2_k} \left\langle\epsilon_k^2 \right\rangle
+\sum_{k k'}  \overline{b_k b_{k'}} \left\langle\epsilon_k^2
\epsilon_{k'}^2\right\rangle =
$$
$$
=\epsilon^2 \overline{a^2} n
+\epsilon^4\sum_{k k'}  \overline{ b_k b_{k'}}
+\sum_{k}  \overline{b_k^2 }\left(
\left\langle\epsilon_k^4\right\rangle -
\left\langle\epsilon_k^2\right\rangle^2
\right)\,,
\eqno(86)
$$
we have for the second cumulant
$$
\left\langle S^2 \right\rangle -
\left\langle S \right\rangle^2 =
\epsilon^2 \overline{a^2} n
+\epsilon^4\sum_{k k'} u_{k k'} +
O\left(\epsilon^4 n \right) \,,
\eqno(87)
$$
where the correlator
$$
u_{k k'}=\overline{ b_k b_{k'}}-\overline{ b_k
}\,\overline{b_{k'}}
\eqno(88)
$$
depends only on the difference $k-k'$, if the distribution of
$b_k$ is stationary; suggesting its exponential falling
on the scale $1/\delta$
$$
u_{k k'}=u_0 e^{-\delta |k-k'|} \,,
\eqno(89)
$$
we have for large  $n$
$$
\left\langle S^2 \right\rangle - \left\langle S
\right\rangle^2 = \overline{a^2} \epsilon^2  n +
u_0\frac{\epsilon^4}{2\delta}  n +
O\left(\epsilon^4 n \right) \,.
\eqno(90)
$$
Therefore, the long-range correlations leads to appearance
of contributions $\epsilon^4/\delta$, which should be
taken into account, in contrast
to corrections of type $\epsilon^4$.

According to (38), the quantities  $a_k$ and $b_k$ are
functions of $\psi_k$. Since
$\left\langle \psi_n \right\rangle=2n\delta$, the phase $\psi$
changes on the scale  $1/\delta$ and
falling of correlations on
the same scale looks rather probable. Absence of corrections
$\epsilon^4/\delta$ for the quantity $z_n$ follows from the
stated above. Existence of such corrections for $\psi_n$ will
be demonstrated below: they are related with approaching the
phase transition, where $\left\langle \psi\right\rangle$
plays a role of the order parameter and turns to zero at the
transition point.

Analogous considerations show that contributions
begin with $\epsilon^4/\delta$ to  the third and fourth
cumulants, with $\epsilon^6/\delta^2$ to the fifth and sixth
cumulants,  and so on.  Hence, for calculation of the first
correction to the main contribution of order $\epsilon^2$ one
should take into account the first four cumulants.

The limit (83) corresponds to a large concentration
of weak scatterers and is usually referred as a "white noise"
potential \cite{603}; near the edge of the initial
band practically any random potential with short-range
correlations reduces to this limit \cite{603}. This limit
is free from effects of commensurability of the Fermi momentum
with the lattice constant \cite{601,602}, which look
hardly observable, but essentially complicate the
mathematical description.

\begin{center}
{\bf 8. The first correction to the Gaussian distribution.  }
\end{center}

According to Sec.4, parameters $\kappa_{m_1 m_2}$ in the
Gaussian approximation are determined by contributions of
order  $\delta$ and $\epsilon^2$. For calculation  the first
correction $\epsilon^4/\delta$ one should take into account the
third and fourth cumulants (Sec.7).

\begin{center}
{\it Allowed band.}
\end{center}

Comparing (37) and (41), we come to conclusion, that in the
allowed band we should retain $S_1$ and $S_4$. Introducing
the characteristic function for the distribution $P(S_1,S_4)$
$$
F(t,t')=\int dS_1 dS_4 P(S_1, S_4) e^{itS_1+it'S_4}\,
\eqno(91)
$$
and writing the expression for the moments
$$
\left\langle y_1^{m_1} y_2^{m_2} \right\rangle
=\int dS_1 dS_4 P(S_1, S_4) e^{i(m_1-m_2)S_1+(m_1+m_2)S_4}\,,
\eqno(92)
$$
one can easily see that it corresponds to the changes
$$
it \to i(m_1\!-\!m_2)\,,\qquad
it' \to (m_1\!+\!m_2)
\eqno(93)
$$
in the definition of the characteristic function. Accepting the
latter in the form (79) and retaining the
third and fourth cumulants, we receive
$$
\kappa_{m_1 m_2} = i\mu_{10}(m_1\!-\!m_2) + \mu_{01}(m_1\!+\!m_2)+
$$
$$
-\mu_{20}(m_1\!-\!m_2)^2+i\mu_{11}(m_1\!-\!m_2)(m_1\!+\!m_2)+
\mu_{02}(m_1\!+\!m_2)^2+
$$
$$
-i\mu_{30}(m_1\!-\!m_2)^3-\mu_{21}(m_1\!-\!m_2)^2(m_1\!+\!m_2)+
$$
$$
+i\mu_{12}(m_1\!-\!m_2)(m_1\!+\!m_2)^2 +\mu_{03}(m_1\!+\!m_2)^3+
$$
$$+
\mu_{40}(m_1\!-\!m_2)^4-i\mu_{31}(m_1\!-\!m_2)^3(m_1\!+\!m_2)-
$$
$$
-\mu_{22}(m_1\!-\!m_2)^2(m_1\!+\!m_2)^2 +
\eqno(94)
$$
$$+
i\mu_{13}(m_1\!-\!m_2)(m_1\!+\!m_2)^3 +\mu_{04}(m_1\!+\!m_2)^4\,.
$$
We have in mind that summation of $n$ analogous terms leads
to the change  $\mu_{ij}\to \mu_{ij} n$, and the exponents
$\kappa_{m_1 m_2}$ are determined by the cumulants $\mu_{ij}$,
corresponding to a single term.

Equation (94) contains 2+3+4+5=14 parameters, which can be
determined using 14 values of $\kappa_{m_1 m_2}$ for the first
four moments. Configurations of $\kappa_{m_1 m_2}$ for even
moments were shown in Fig.4, while for odd moments they are
presented in Fig.5. The evolution equations for odd moments
are derived in Appendix 2, where expansions of $\kappa_{m_1 m_2}$
in $\epsilon^2/\delta$ are also given.

\begin{figure}
\centerline{\includegraphics[width=2.6 in]{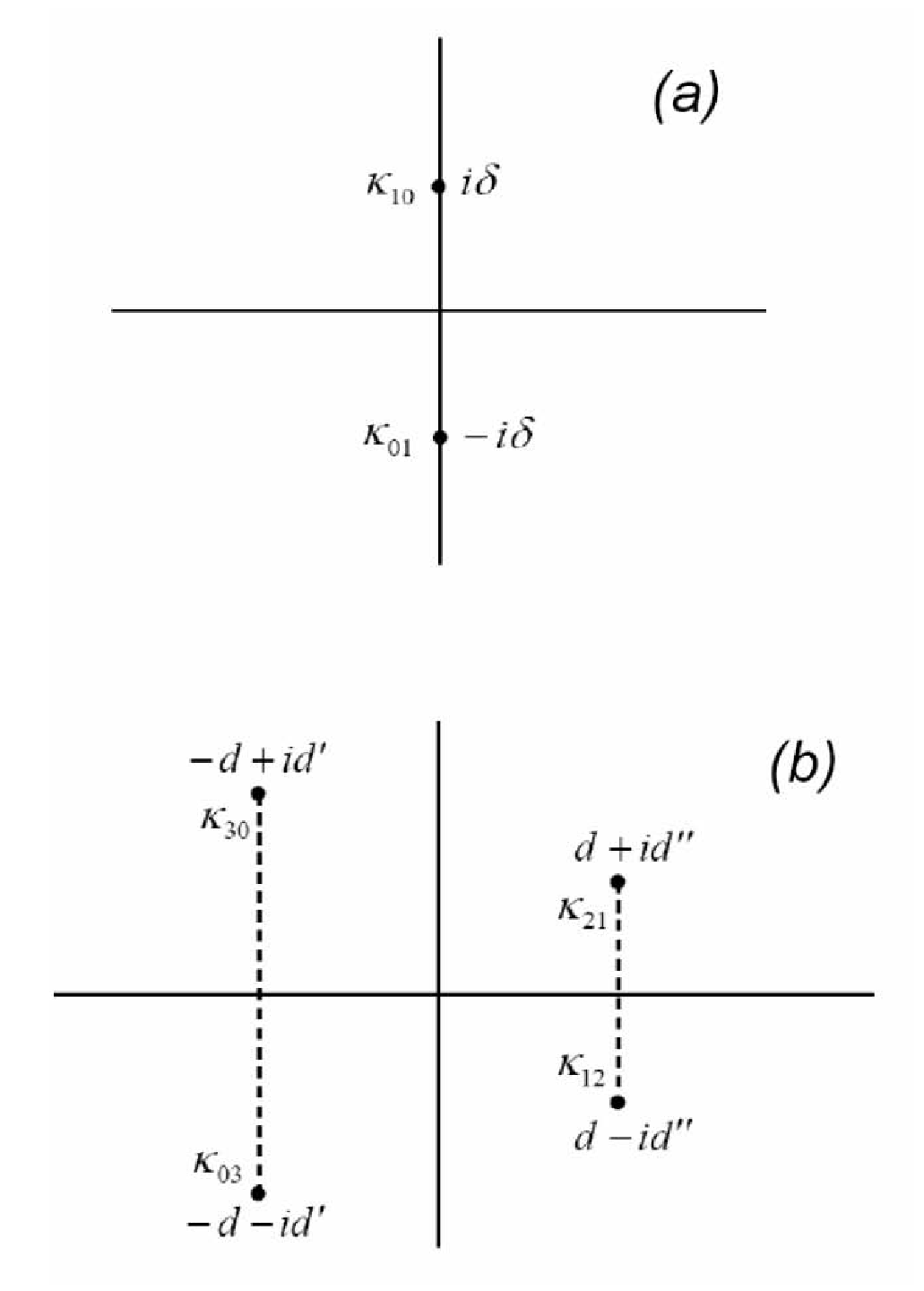}}
\caption{ \small Configuration of the exponents
$\kappa_{m_1 m_2}$ for the first and third moments in the
depth of the allowed band.  } \label{fig5} \end{figure}

In the allowed band the exponents $\kappa_{m_1 m_2}$
and $\kappa_{m_2 m_1}$ are complex-conjugated; their real
and imaginary parts  are determined by symmetric and
antisymmetric combinations (47) correspondingly. Corrections of
order $\epsilon^4/\delta$ take place only for antisymmetric
combinations (see Appendix 2), while for symmetric ones
they begin from $\epsilon^6/\delta^2$, being absent in the
accepted approximation. As a result, parameters $\mu_{ij}$ for
the symmetric combinations coincide with their Gaussian values
$$
\mu_{01}=\epsilon^2/2 \,,\quad
\mu_{20}=3\epsilon^2/4 \,,\quad
\mu_{02}=\epsilon^2/4 \,,\quad
$$
$$
\mu_{21}=\mu_{03}=\mu_{40}=\mu_{22}=\mu_{04}=0\,.
\eqno(95)
$$
Considering antisymmetric combinations  and comparing with
Figs.4,5, we have a system of equations for $\mu_{10}$, $\mu_{11}$
$\mu_{30}$, $\mu_{12}$, $\mu_{31}$, $\mu_{13}$. Solution of this
system with use of  the specific values of parameters  (see
Appendix 2)
$$
a'=2\delta+\frac{3}{4} \frac{\epsilon^4}{\delta}\,,\quad
b'=4\delta+\frac{33}{2} \frac{\epsilon^4}{\delta}\,,\quad
c'=2\delta-\frac{69}{4} \frac{\epsilon^4}{\delta}\,,\quad
$$
$$
d'=3\delta+\frac{21}{4} \frac{\epsilon^4}{\delta}\,,\quad
d''=\delta-\frac{27}{4} \frac{\epsilon^4}{\delta}\,,\quad
\eqno(96)
$$
leads to results
$$
\mu_{10}=\delta+\frac{5}{8} \frac{\epsilon^4}{\delta}\,,\quad
\mu_{11}=-\frac{9}{8} \frac{\epsilon^4}{\delta}\,,\quad
\eqno(97)
$$
$$
\mu_{30}=-\frac{17}{16} \frac{\epsilon^4}{\delta}\,,\quad
\mu_{12}=-\frac{9}{16} \frac{\epsilon^4}{\delta}\,,\quad
\mu_{31}=0\,,\quad \mu_{13}=0\,.
$$
Disappearance of  $\mu_{31}$ and $\mu_{13}$ corresponds to
complete neglect of fourth cumulants, i.e. they are
absent in the order $\epsilon^4/\delta$. The situation will
be different in the forbidden band
(see below).\,\footnote{\,There is similarity with papers
\cite{403,404}  that the character of deviation from the Gaussian
distribution signals of appearance of the phase transition. }

\begin{center}
{\it Forbidden band.}
\end{center}

In the forbidden band solutions $y_1$ and $y_2$ are real
(Sec.5), but can change their sign:  it leads to contributions
$\pm i\pi$ in the exponentials of (41). If $y_1$ and $y_2$ change
sign in different points, then both $S_1$ and $S_3$ in Eq.41
become finite. Since $y_1^2$ and $y_2^2$ are real
and positive, one can calculate the moments with even $m_1$
and $m_2$ neglecting $S_1$ and $S_3$.
Introducing the characteristic function for the distribution
$P(S_2,S_4)$ analogously (91), it is easy to establish that
calculation of averages $\left\langle y_1^{m_1} y_2^{m_2} \right\rangle$
corresponds to the change
$$
it \to (m_1\!-\!m_2)\,, \qquad it' \to (m_1\!+\!m_2)
\eqno(98)
$$
in the definition of the characteristic function. Accepting the
latter in the form  (79), we have for the exponents
$\kappa_{m_1 m_2}$ expression analogous to (94): this result is
valid for even $m_1$ and $m_2$. For other moments one should take
$S_1$ and $S_3$ into account.  Let introduce the characteristic
function
$$
F(t,t',t'',t''')=\int dS_1 dS_2 dS_4 dS_3 P(S_1,S_2,S_4, S_3)
\cdot
$$
$$
\cdot
e^{itS_1+it'S_2+it''S_4+it'''S_3}  \,,
\eqno(99)
$$
where we have changed succession of $S_i$ in $P\{S_i\}$:
then the first and last indices of $\mu_{k k' k''k'''}$
in the expression
$$
F(t,t',t'',t''')=\exp  \sum_{k k' k'' k'''} \mu_{k k' k''
k'''} (it)^{k}(it')^{k'}\cdot
$$
$$ \cdot(it'')^{k''} (it''')^{k'''} \, \eqno(100) $$ correspond
to imaginary contributions related with $S_1$ and $S_3$,  which
are added to real contributions, arising from $S_2$ and $S_4$
and describing by the middle indices. Composing
averages $\left\langle y_1^{m_1} y_2^{m_2} \right\rangle$,
one can see that they correspond to the changes in Eq.99
$$
it \to i(m_1\!-\!m_2)\,, \qquad it' \to (m_1\!-\!m_2)
\,,
$$
$$ it'' \to (m_1\!+\!m_2)\,,
\qquad it''' \to i(m_1\!+\!m_2)
\eqno(101)
$$
so that
$$
\left\langle y_1^{m_1} y_2^{m_2} \right\rangle
=\exp \sum_{k k' k'' k'''} \mu_{k k' k'' k'''}
[i(m_1\!-\!m_2)]^{k} \cdot
$$
$$
\cdot (m_1\!-\!m_2)^{k'}(m_1\!+\!m_2)^{k''}
[i(m_1\!+\!m_2)]^{k'''} \,. \eqno(102)
$$
In the deep of the forbidden band all $\kappa_{m_1m_2}$
are real, which leads to vanishing of $\mu_{k k'
k'' k'''}$ with odd $k+k'''$.
The obtained expression for $\kappa_{m_1 m_2}$ has the
structure analogous (94), but with the renormalized values
$\tilde\mu_{ik}$ instead of $\mu_{ik}$:
$$
\tilde\mu_{10}=\mu_{10}\,,\quad \tilde\mu_{01}=\mu_{01}\,\quad
\tilde\mu_{20}=\mu_{20}-\mu_{2000}\,,
$$
$$
\tilde\mu_{11}=\mu_{11}-\mu_{1001}\,,\quad
\tilde\mu_{02}=\mu_{02}-\mu_{0002}\,,\quad
$$
$$
\tilde\mu_{30}=\mu_{30}-\mu_{2100}\,,\quad
\tilde\mu_{21}=\mu_{21}-\mu_{2010}-\mu_{1101}\,,\quad
$$
$$
\tilde\mu_{12}=\mu_{12}-\mu_{0102}-\mu_{1011}\,,\quad
\tilde\mu_{03}=\mu_{03}-\mu_{0012}\,,\quad
$$
$$
\tilde\mu_{40}=\mu_{40}+\mu_{4000}-\mu_{2200}\,,\quad
$$
$$
\tilde\mu_{31}=\mu_{31}+\mu_{3001}-\mu_{2110}-\mu_{1201}\,,\quad
\eqno(103)
$$
$$
\tilde\mu_{22}=\mu_{22}+\mu_{2002}-\mu_{2020}-\mu_{0202}
-\mu_{1111}\,,\quad
$$
$$
\tilde\mu_{13}=\mu_{13}+\mu_{1003}-\mu_{1021}-\mu_{0112}\,,\quad
$$
$$
\tilde\mu_{04}=\mu_{04}+\mu_{0004}-\mu_{0022}\,.
$$
For coefficients of symmetrical combinations corrections
$\epsilon^4/\delta$ are absent independently of
parity of $m_1$ and $m_2$, and they retain their Gaussian
values
$$
\tilde\mu_{01}=\mu_{01}=-\epsilon^2/2 \,,\quad
\tilde\mu_{20}=\mu_{20}=3\epsilon^2/4 \,,\quad
$$
$$
\tilde\mu_{02}=\mu_{02}=-\epsilon^2/4 \,,\quad
\tilde\mu_{21}=\mu_{21}=0\,,\quad
$$
$$
\tilde\mu_{03}=\mu_{03}=0\,,\quad
\tilde\mu_{40}=\mu_{40}=0\,,\quad
$$
$$
\tilde\mu_{22}=\mu_{22}=0\,,\quad
\tilde\mu_{04}=\mu_{04}=0\,.
\eqno(104)
$$
Antisymmetric combinations and parametrization of
exponents $\kappa_{m_1m_2}$ corresponds to Eq.47 and Figs.4,5
without factors $i$, while the parameters
$a'$, $b'$, $c'$, $d'$, $d''$ are determined by
expressions (96) with different signs of corrections
$\epsilon^4/\delta$. For the moments with even $m_1$ and $m_2$
the required conditions are formulated in terms of $\mu_{ij}$
$$
\tilde\kappa_{20}=2\mu_{10} +4\mu_{11}+8(\mu_{30}\!+\!\mu_{12})
               +16(\mu_{31}\!+\!\mu_{13})=a'\,,
$$
$$
\tilde\kappa_{40}=4\mu_{10} +16\mu_{11}+64(\mu_{30}\!+\!\mu_{12})
               +256(\mu_{31}\!+\!\mu_{13})=b'\,,
\eqno(105)
$$
so there are two conditions for four combinations. Additional
conditions can be found, if variables $z_1$ and $z_2$ are
introduced by relations (63);
then one can find that the distribution
$P(z_1,z_2)$ corresponds to the
characteristic function $F(t_1,t_2)$ with $t_1=(t'+t'')/2$,
$t_2=(t''-t')/2$. To obtain the characteristic function for
the distribution $P(z_1)$, one should set $t=0$, $t'=t''$,
$t'''=0$ in $F(t,t',t'', t''')$; then
$$
\ln F_{z_1}(t)= (\mu_{10}+\mu_{01})(it)+
(\mu_{20}+\mu_{11}+\mu_{02})(it)^2+
$$
$$
+(\mu_{30}+\mu_{21}+\mu_{12}+ \mu_{03})(it)^3+
$$
$$
+(\mu_{40}+\mu_{31}+\mu_{22}+\mu_{13} +\mu_{04})(it)^4
\eqno(106)
$$
However, the distribution of $z_1$ is Gaussian in the limit (83)
(Sec.7), so contributions $(it)^3$ and $(it)^4$ should be
absent; using vanishing of the coefficients for symmetrical
combinations, we have two conditions
$$
\mu_{30}+\mu_{12}=0\,, \quad
\mu_{31}+\mu_{13} =0 \,,
\eqno(107)
$$
which allow to resolve (105) for parameters
$$
\mu_{10}= \frac{4a'-b'}{4}= \delta+\frac{27}{8}
\frac{\epsilon^4}{\delta}\,,\quad
\mu_{11}= \frac{b'-2a'}{8}=-\frac{15}{8}
\frac{\epsilon^4}{\delta}\,.
\eqno(108)
$$
Expressions (108) reproduce the correct values
of $v$ and $D$ in the forbidden band with required accuracy.

For the rest of moments, the conditions are formulated
in terms of $\tilde\mu_{ij}$ and give four equations
for six quantities $\tilde\mu_{10}$, $\tilde\mu_{11}$
 $\tilde\mu_{30}$, $\tilde\mu_{12}$, $\tilde\mu_{31}$,
 $\tilde\mu_{13}$.
With the use of equality $\tilde\mu_{10}=\mu_{10}$ (see Eq.103)
it leads to relations
$$
\tilde\mu_{11}=\mu_{11}+\delta\mu_{11}\,,\quad
\tilde\mu_{30}+\tilde\mu_{12}
=-\frac{41}{18} \frac{\epsilon^4}{\delta}
-\frac{4}{3} \delta \mu_{11} \,,\quad
$$
$$
\tilde\mu_{31}+\tilde\mu_{13}
=\frac{7}{9}\frac{\epsilon^4}{\delta}
+\frac{1}{3} \delta \mu_{11}   \,,\quad
\eqno(109)
$$
$$
\tilde\mu_{30}+3\tilde\mu_{31}=
-\frac{17}{16} \frac{\epsilon^4}{\delta}\,,\qquad
\tilde\mu_{30}+4\tilde\mu_{31}=
-\frac{7}{144} \frac{\epsilon^4}{\delta}
+\frac{1}{3} \delta \mu_{11}\,,\quad
$$
and comparison with (107) shows that equality
$\tilde\mu_{ij}=\mu_{ij}$ for arbitrary  $i,j$  cannot be
reached, independently of the value of $\delta
\mu_{11}$:  it is a direct evidence of finiteness of $S_1$ and
$S_3$.

The characteristic functions for distributions $P(S_1)$ and
$P(S_3)$ are obtained from (100) at $t'=t''=t'''=0$ and
$t=t'=t''=0$ correspondingly: it eliminates all
coefficients apart from $\mu_{k000}$ or $\mu_{000k}$.
Comparison of (103) and (104) shows vanishing of $\mu_{2000}$
and $\mu_{0002}$, and we are left with characteristic functions
$$
F_{S_1}(t)=\exp\left\{ \mu_{4000} (it)^4 \right\}\,,\quad
F_{S_3}(t)=\exp\left\{ \mu_{0004} (it)^4 \right\}\,.
\eqno(110)
$$
In both cases we have the distribution
$$
P(x)= \frac{1}{2\pi} \int dt  e^{-ixt -a t^4}=
 \frac{1}{a^{1/4}} \bar P\left(\frac{x}{a^{1/4}}\right)\,,
 \eqno(111)
$$
which is well-defined for  $a>0$; here $\bar P\left(x \right)$
corresponds to $a=1$. The distribution $P(x)$ is even in $x$,
so $\left\langle S_1 \right\rangle= \left\langle S_3
\right\rangle=0$. If sums $S_1$ and $S_3$ contains $n$ terms
then $a$ has a linear growth in $n$, $a=\bar a n$, and the width
of the distribution extends as $n^{1/4}$ instead of $n^{1/2}$
for the Gaussian case.

\begin{center}
{\bf 9. Complete distribution for ${\cal E}>{\cal E}_0$ }
\end{center}

The first two terms in Eq.37 should be complex-conjugated,
and for the proper choice of the $x$ origin can be written in the
form
$$
y_1=   e^{z/2-i\psi/2} \cos{(\chi/2)}\,, \quad
y_2=  e^{z/2+i\psi/2} \cos{(\chi/2)} \,.
\eqno(112)
$$
If $\cos(\chi/2)$
is carried to the exponential, then a term $\ln\cos(\chi/2)$
appears in it; a real part of this term is of no
interest due to absence of the systematic growth of the phase
$\chi$ \cite{125}. Its imaginary part is absent for
$\cos(\chi/2)>0$, and reduces to $\pm i\pi$ in the opposite
case. For the proper choice of the sign one can write
$$
y_1=   e^{z/2-i\psi/2-if(\chi)} \,, \quad
y_2=  e^{z/2+i\psi/2 +if(\chi)}  \,,
\eqno(113)
$$
where $f(\chi)={\rm Im}\left\{\ln\cos(\chi/2)\right\}$.
As a result, only quantities $S_1$ and $S_4$ remain in
Eq.44, whose mutual distribution is defined as
$$
P\left( S_1, S_4 \right) =\int
\delta\left( S_1\!-\!\psi/2\!-\!f(\chi) \vphantom{S_1^2} \right)
\delta\left(S_4\!-\!z/2  \vphantom{S_1^2}\right) \cdot
$$
$$ \cdot
P\left(z,\psi,\chi\right) dz d\psi
d\chi \eqno(114)
$$
and leads to the characteristic function
$$
F \left( t, t' \right) =\int e^{it\psi/2+itf(\chi)+it'z/2}
 P\left(z,\psi,\chi\right) dz d\psi d\chi \,.
\eqno(115)
$$
The moments $\left\langle y_1^{m_1} y_2^{m_2} \right\rangle$
of our interest are obtained by the changes (93).
The distribution function $P\left(z,\psi,\chi\right)$
follows from the distribution  $P\left(\rho,\psi,\chi\right)$
studied previously \cite{125} in the result of substitution (33).
As a result, we have the formal solution for the
distribution $P(y_1, y_2)$,  but its practical application
needs a large-scale numerical work.

\begin{center}
{\bf 10. Broadening of spectral lines in the universal
conductance fluctuations. } \end{center}

In the deep of the allowed band all exponents $\kappa_{m_1 m_2}$
with $m_1\ne m_2$ are complex-valued. Their imaginary parts
determine discrete frequencies for oscillations of moments
$\left\langle \rho^n \right\rangle$ of the
Landauer resistance $\rho$: these frequencies can be revealed
by the spectral analysis of the universal conductance
fluctuations \cite{122,123}. Indeed, the characteristic function
$F(t)$, corresponding to the distribution $P(\rho)$, is
the generating functions of moments
$\left\langle \rho^n \right\rangle$  (compare with (73)); if
the latter are known, one can construct the function $F(t)$ and
find the distribution $P(\rho)$ by the inverse Fourier transform.
Due to oscillations of moments $\left\langle \rho^n
\right\rangle$, the distribution $P(\rho)$ is represented as a
superposition of discrete harmonics: it leads to aperiodic
oscillations of $\rho$ in the given sample.

Averages $\left\langle \rho^n \right\rangle$ are related
with  even moments of solutions $y_i$, and  for determination
of discrete frequencies one should set
$$
m_1-m_2=2k\,, \quad m_1+m_2=2n\, \quad (n\ge k)
\eqno(116)
$$
in the antisymmetric combinations $\tilde\kappa_{m_1m_2}$,
corresponding to Eq.94. Due to vanishing of $\mu_{31}$ and
$\mu_{13}$ (see Eq.97), we have the following values for the
discrete frequencies
$$
\omega_{n,k}=2k\left[  \mu_{10}+2\mu_{11}n -4\mu_{30} k^2 +
4 \mu_{12} n^2\right]\,,
$$
$$ n=k,k+1,k+2,\ldots
\eqno(117)
$$
In the extremely metallic limit we can set $\mu_{10}=\delta$,
$\mu_{11}=\mu_{30}= \mu_{12}=0$ and obtain frequencies
$\omega_{n,k}=2k\delta$ with evident degeneracy in $n$.
If corrections $\epsilon^4/\delta$ are taken into account,
this degeneracy is removed, and instead of a single line with
the fixed $k$ value  the  set of the satellite lines
arises. Their frequencies can be obtained, if $n$ runs from
$k$ to infinity:  for small $\epsilon^4/\delta$ it looks as a
broadening of the initial degenerate line. The effective
intensity of the satellite lines decreases with $n$, and
can be estimated in the following manner.

In the depth of the allowed band one can use the random phase
approximation, which gives the following distribution $P(\rho)$
for the small system length $L$ \cite{10,12}
$$
P(\rho)=\frac{e^{-\rho/\alpha L} }
{\alpha L}\,.
\eqno(118)
$$
The contribution of the $n$th moment to oscillations is
determined by the quantity
$$
\frac{\left\langle \rho^n \right\rangle }{n!}
=\left( \alpha L\right)^n\,.
\eqno(128)
$$
The oscillations are suppressed for $\alpha L \agt 1$  due
to transition to the log-normal distribution. For small
$\alpha L$  one has strong suppression of higher moments
and the corresponding higher harmonics. The well-developed
picture of aperiodic oscillations, corresponding to universal
conductance fluctuations, is realized for $\alpha L\sim 1$:
for example, the spectral analysis \cite{122,123} of the
classical results by Webb and Washburn \cite{305} reveals
7 harmonics of essential amplitude, which corresponds to the
estimate $\alpha L\approx 0.85$ for an average point of the
actual interval of lengths. Thus, for illustration of
the broadening of spectral lines (Fig.6) we accept the estimate
$$
A_{n,k}\sim 0.85^{\,n}
\eqno(129)
$$
for the amplitudes of the satellite lines. One can see that
broadening is essential even for a rather small value
$\epsilon^4/\delta^2=0.0005$.

\begin{figure}
\centerline{\includegraphics[width=2.4 in]{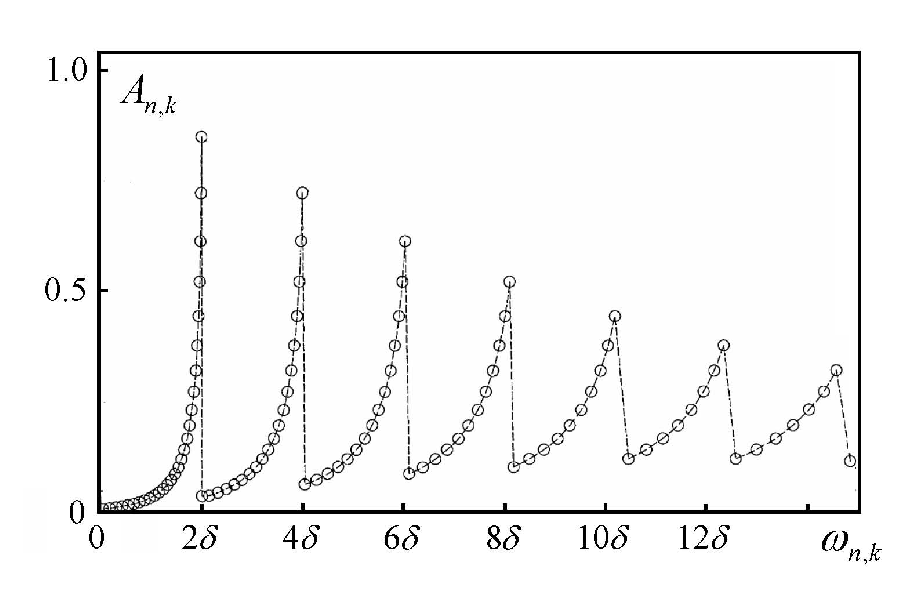}}
\caption
{ \small Broadening of spectral lines in the universal
conductance fluctuations,
related with removing of degeneracy in $n$,
for $\epsilon^4/\delta^2=0.0005$. The estimate (129)
is accepted for the amplitudes of the satellite lines.}
\label{fig6} \end{figure}

\begin{center}
{\bf 11. Conclusion }
\end{center}

In the present work we considered the mutual distribution of
two linearly independent solutions $y_1(x)$ and $y_2(x)$
of the 1D Schr${\rm\ddot o}$dinger equation with a random
potential, which determine evolution of moments
$\left\langle y_1^{m_1}y_2^{m_2} \right\rangle$. In terms of the
logarithmic variables $z_1=\ln y_1$, $z_2=\ln y_2$ the
distribution is Gaussian in the deep of the allowed and forbidden
bands.  Corrections to the Gaussian distribution can be
calculated in the form of expansion in the parameter
$\epsilon^2/\delta$ by  successive inclusion of the higher
cumulants. The first correction to the Gaussian distribution is
calculated explicitly, and used for the demonstration of
broadening of the spectral lines in the universal conductance
fluctuations \cite{122,123}. The complete distribution of $y_1$
and $y_2$ is formally expressed in terms of the distribution
$P(\rho,\psi,\chi)$, discussed in  \cite{125}.

New information is obtained on the phase transition in
the distribution $P(\psi)$, predicted previously
\cite{124,125,126}.  The real sense of the transition consists in
the change of a configuration of four Lyapunov exponents, which
determine the general solution: instead of two pairs of the
complex-conjugated exponents, four real exponents appear (Fig.3).
Such change of the regime is confirmed by results in the deep of
allowed and forbidden bands: it proves the existence of the
singular point ${\cal E}_0$ at the formal level. At the point
${\cal E}_0$ the phase $\psi$ acquires an imaginary part, while
its real part vanishes, and redefinition occurs of the
Lyapunov exponents and the Landauer resistance.
Also, the relationship breaks
between the Landauer resistance $\rho$ and the
moments of the Cauchy solution: in the allowed band
$\rho$ is determined by combinations $\left\langle y_1^{m} y_2^{m}
\right\rangle$, while in the forbidden band it is related with
averages $\left\langle y_1^{2m} y_2^{0} \right\rangle$.

%

The transfer matrix approach to quasi-1D systems gives
the popular numerical algorithm \cite{131} for
estimation of the critical properties for the Anderson
transition (see references in \cite{200,129}). It is based on
the scaling hypothesis that dependence on the transverse size
$L$ of the system is described by a function $F(L/\xi)$, where
$\xi$ is the localization length. The behavior of the second
moments of the Cauchy solution in quasi-1D systems is calculated
comparatively simply \cite{129,130}, but applying the
scaling hypothesis to it leads to contradiction with numerical
results \cite{130} and indicates the necessity for the more subtle
analysis. The change of relation between the moments and Lyapunov
exponents, discovered in the present paper, looks rather
fundamental and having a great significance for the Anderson
transition theory.

%
%
%
%
%
%
%
%
%


\begin{center}
{\it Appendix 1.}{\bf \,\,
Derivation of the result for  $\kappa_{m_1
m_2}$} \end{center}

The exponential average, which is of interest for us, is
defined as
$$\!\!\!\!
\left\langle  y_1^{m_1} y_2^{m_2}   \right\rangle \!=\!\!
\int \prod_i dS_i \,
{\rm e}^{(m_1-m_2)(iS_1+S_2)+(m_1+m_2)(iS_3+S_4)} \cdot
$$
$$ \cdot
\exp \left\{ -\frac{1}{n} \sum_{ij} B_{ij}
\left( S_i-v_i n \right) \left( S_j-v_j n \right) \right\},
\eqno(A.1)
$$
if the normalization is temporally ignored. Let accept
$$
S_i=v_in+x_i+\Delta_i
\eqno(A.2)
$$
and choose $\Delta_i$ so that to eliminate
the linear in $x_i$ terms in the exponential. Then
\onecolumn

$$
\left\langle  y_1^{m_1} y_2^{m_2}   \right\rangle =
{\rm e}^{(m_1-m_2)(iv_1 n+v_2 n+i\Delta_1 +\Delta_2)}\cdot
$$
$$ \cdot\,
{\rm e}^{(m_1+m_2)(iv_3 n+v_4 n+i\Delta_3 +\Delta_4) }
\exp \left\{ -\frac{1}{n} \sum_{ij} B_{ij} \Delta_i \Delta_j
\right\}
\int \prod_i dx_i \,
\exp \left\{ -\frac{1}{n} \sum_{ij} B_{ij} x_i x_j \right\} .
\eqno(A.3)
$$
Setting
$$
C_i = \sum_j 2B_{ij} \Delta_j
\eqno(A.4)
$$
we obtain the following conditions for vanishing of the
linear terms
$$
C_1=i(m_1-m_2)n\,,\quad
C_2=(m_1-m_2)n\,,\quad
C_3=i(m_1+m_2)n\,,\quad
C_4=(m_1+m_2)n\,.
\eqno(A.5)
$$
Introducing the matrix $||A_{ij}||$, inverse to
$||B_{ij}||$, we have
$$
2 \Delta_i=  \sum_j A_{ij} C_j
\eqno(A.6)
$$
or more specifically
$$
\Delta_k= \frac{n}{2} \left[ (m_1-m_2) (iA_{k1}+A_{k2})+
+(m_1+m_2) (iA_{k3}+A_{k4})\right]  \,.
\eqno(A.7)
$$


\noindent
Substituting to $(A.3)$, using relation
$$
 \sum_{kl} B_{kl} A_{ki}A_{lj} =
 \sum_{k} A_{ki}\delta_{kj} = A_{ji}
\eqno(A.8)
$$
for simplification of combinations with $\Delta_i$,
and removing the integral over $x_i$ by the normalization
condition ($\left\langle  y_1^0 y_2^0
\right\rangle = 1$), we come to result (54) with the exponents
$\kappa_{m_1 m_2}$, determined by (55).


\begin{center}
{\it Appendix 2.}{\bf \,\,Evolution of odd moments
and expansions for $\kappa_{m_1 m_2}$ }
\end{center}

According to \cite{13}, the evolution of moments is
conveniently considered for the forbidden band, while
 description of the allowed band is obtained as analytical
continuation. Evolution of the elements $t_{ij}$
of the pseudo transfer matrix is described by equations
$$
x_{n}= u_n x_{n-1} + \tilde v_n y_{n-1}\,,
\quad
y_{n}= v_n x_{n-1} + \tilde u_n y_{n-1} \,,
\eqno(A.9)
$$


\noindent
where $x_n$, $y_n$ correspond to the pair $t_{11}^{(n)}$,
$t_{12}^{(n)}$, or  $t_{21}^{(n)}$,
$t_{22}^{(n)}$. Here  $x_{n-1}$, $y_{n-1}$ are
statistically independent of $\epsilon_n$, and
$$
u_n=(1+\epsilon_n) e^{-\delta}\,, \quad v_n=\epsilon_n
e^{-\delta}  \,,\qquad\,\,
\tilde u_n=(1-\epsilon_n) e^{\delta}\,,
\quad \tilde v_n=-\epsilon_n
e^{\delta}   \,.
\eqno(A.10)
$$


Introducing notations for the third moments
$$
z_{1}^{(n)}=\left\langle x_n^{3}\right\rangle\,, \quad
z_{2}^{(n)}=\left\langle x_n^{2} y_n\right\rangle\,,\quad
z_{3}^{(n)}=\left\langle x_n y_n^{2}\right\rangle\,, \quad
z_{4}^{(n)}=\left\langle y_n^{3}\right\rangle\,,
\eqno(A.11)
$$
we come to the system of the linear difference equations
with constant coefficients


$$
\left ( \begin{array}{cccc} z_{1}^{(n)} \\ z_{2}^{(n)}
\\ z_{3}^{(n)} \\ z_{4}^{(n)}\end{array} \right)\,
= \left ( \begin{array}{cccc} 1-3\delta +3\epsilon^2 &
-6\epsilon^2 & 3\epsilon^2 & 0
\\ 2\epsilon^2 & 1-\delta-3\epsilon^2 & 0 & \epsilon^2\\
\epsilon^2 & 0 & 1+\delta-3 \epsilon^2 & 2\epsilon^2 \\
0 & 3\epsilon^2 & -6\epsilon^2 & 1+3\delta+ 3\epsilon^2
 \end{array} \right)\,
\left ( \begin{array}{ccc} z_{1}^{(n-1)} \\ z_{2}^{(n-1)}
\\z_{3}^{(n-1)} \\z_{4}^{(n-1)} \end{array} \right) \,,
\eqno(A.12)
$$
whose solution is searched in the exponential form, $z_{i}^{(n)} \sim
\lambda^n $ \cite{30}; it is easy to see that $\lambda$ is an
eigenvalue of the matrix  $(A.12)$. Setting
$\lambda=1+\kappa$, we have the equation for determination
of  $\kappa$
$$
\left(\kappa^2-9\delta^2\right) \left(\kappa^2-\delta^2\right)
= 48\epsilon^2 \delta^2  \kappa\,,
\eqno(A.13)
$$
\twocolumn

\noindent
whose roots in the depth of the forbidden band allow
the asymptotic expansions
$$
\kappa_{30}=3\delta+3\epsilon^2-\frac{21}{4}
           \frac{\epsilon^4}{\delta}\,,\quad
\kappa_{21}=\delta-3\epsilon^2+\frac{27}{4}
           \frac{\epsilon^4}{\delta}\,,\quad
$$
$$
\kappa_{03}=-3\delta+3\epsilon^2+\frac{21}{4}
           \frac{\epsilon^4}{\delta}\,,\quad
\kappa_{12}=-\delta-3\epsilon^2-\frac{27}{4}
           \frac{\epsilon^4}{\delta}\,.
\eqno(A.14)
$$
Analogous equations for the second and the fourth
moments were obtained previously \cite{13}
$$
\kappa\left(\kappa^2-4\delta^2\right)=
8\epsilon^2 \delta^2  \,,
\eqno(A.15)
$$
$$
\kappa\left(\kappa^2-4\delta^2\right)\left(\kappa^2-16\delta^2\right)=
24\epsilon^2 \delta^2 \left(7 \kappa^2-16 \delta^2\right ) \,.
\eqno(A.16)
$$
In the depth of the forbidden band their roots allow the
expansions
$$
\kappa_{20}=2\delta+\epsilon^2-\frac{3}{4}
           \frac{\epsilon^4}{\delta}\,,\quad
\kappa_{11}=-2\epsilon^2+O\left(\frac{\epsilon^6}{\delta^2}
\right) \,,\quad
$$
$$
\kappa_{02}=-2\delta+\epsilon^2+\frac{3}{4}
           \frac{\epsilon^4}{\delta}\,,\quad
\eqno(A.17)
$$

$$
\kappa_{40}=4\delta+6\epsilon^2-\frac{33}{2}
           \frac{\epsilon^4}{\delta}\,,\quad
\kappa_{31}=2\delta-3\epsilon^2+\frac{69}{4}
           \frac{\epsilon^4}{\delta}\,,\quad
 $$
 $$
 \kappa_{22}=-6\epsilon^2+O\left(\frac{\epsilon^6}{\delta^2}
\right) \,,\quad
\eqno(A.18)$$
$$
\kappa_{04}=-4\delta+6\epsilon^2+\frac{33}{2}
           \frac{\epsilon^4}{\delta}\,,\quad
\kappa_{13}=-2\delta-3\epsilon^2-\frac{69}{4}
           \frac{\epsilon^4}{\delta}\,.\quad
$$
One can observe, that for calculation of the exponents
$\kappa_{m_1 m_2}$ with accuracy $\epsilon^2$, i.e. in the lowest
order in a random potential, it is sufficient to retain
only diagonal elements of the matrix $(A.12)$, since
contribution of non-diagonal elements begins with $\epsilon^4$.
Such "diagonal" approximation can be realized for the
moments of the arbitrary order \cite{123}, and leads to the
results (27) and (53).


For the first moments we have the trivial equation
$$
\left(\kappa^2-\delta^2\right)= 0 \,,
\eqno(A.19)
$$
not containing a random potential. In all equations
$(A.13)$, $(A.15)$, $(A.16)$, $(A.19)$ we observe
vanishing of the coefficient for the next to leading power of
$\kappa$, so the sum of the roots turns to zero due to
the generalized Viete theorem.

For the 1D Anderson model we have $\delta^2=-{\cal E}$,
$4\epsilon^2\delta^2=W^2$, where ${\cal E}$ is the energy counted
from the lower edge of the initial band, $W$ is the amplitude
of the random potential. Equations  $(A.13)$, $(A.15)$, $(A.16)$,
$(A.19)$ were derived for ${\cal E}<0$, but can be analytically
continued to positive ${\cal E}$ due to their regularity in
energy. The results for the allowed band are obtained by the
changes $\delta\to i\delta$, $\epsilon\to i\epsilon$ in all
previous expressions.


\end{document}